\documentclass[letterpaper,twocolumn,10pt]{article}
\usepackage{usenix2024_SOUPS}

\usepackage{tikz}
\usepackage{amsmath}
\usepackage{multirow}
\usepackage[most]{tcolorbox}

\usepackage{xltabular}
\usepackage{booktabs}
\usepackage{amssymb}  % for \bigcirc

\begin{document}
%-------------------------------------------------------------------------------
% show page number
% \pagestyle{empty}

%don't want date printed
\date{}

\title{\Large \bf Is It Real? Exploiting Virtual-Physical Discrimination \\Vulnerability in Mixed Reality}

% if you leave this blank it will default to a possibly ugly attempt 
% to make the contents of the \author command below into a string
\def\plainauthor{Author name(s) for PDF metadata. Don't forget to anonymize for submission!}

%for single author (just remove % characters)
\author{
{\rm Xueyang Wang}\\
Tsinghua University
\and
{\rm Xihuan Yao}\\
Tsinghua University
\and
{\rm Yanming Xiu}\\
Duke University
\and
{\rm Xin Yi\thanks{Corresponding author. Email: yixin@tsinghua.edu.cn.}}\\
Tsinghua University\\
Beijing Academy of Artificial Intelligence
\and
{\rm Maria Gorlatova}\\
Duke University
\and
{\rm Hewu Li}\\
Tsinghua University
} % end author

\maketitle
\thecopyright

%-------------------------------------------------------------------------------
\begin{abstract}
%-------------------------------------------------------------------------------
Consumer mixed reality (MR) headsets seamlessly blend virtual content into physical environments with sufficient fidelity that users may be unable to distinguish virtual objects from physical ones. We identify this virtual-physical discrimination vulnerability as an exploitable security primitive. Through speculative design workshops with 12 experts from cybersecurity and MR/HCI, we develop a taxonomy of virtual-physical confusion attacks and implement four proof-of-concept attacks on Apple Vision Pro, evaluating them with 26 participants in realistic MR tasks. All four attacks altered user behavior, with success rates ranging from 85\% to 100\%, producing misdirected interactions, misjudged object identities, biased purchasing decisions, and altered navigation paths. Notably, the most successful attacks were also the hardest to detect according to participants' subjective ratings. Even participants who recognized virtual content still complied behaviorally, and no participant attributed anomalous events to adversarial causes. We propose platform-level provenance, interaction gating, and user education as countermeasures.
\end{abstract}

%-------------------------------------------------------------------------------
\section{Introduction}
%-------------------------------------------------------------------------------

Mixed reality (MR) systems merge virtual content with the physical world, enabling users to interact with digital objects situated in their real surroundings. Consumer devices such as Apple Vision Pro and Meta Quest 3 now achieve high visual fidelity through advanced rendering, real-time environment understanding, and precise spatial anchoring~\cite{itoh2021towards}, underpinning a growing ecosystem of task-assistance applications where virtual content blends seamlessly into physical environments~\cite{wang2020comprehensive, auda2023scoping, speicher2019mixed}.

This seamless blending introduces a fundamental perceptual challenge~\cite{drascic1996perceptual, kruijff2010perceptual}. When virtual objects are rendered with sufficient realism and contextual consistency, users cannot reliably determine which objects are physically present and which are digitally generated. In non-adversarial settings, users have already sat on virtual chairs without physical verification~\cite{wiesing2025confusing}, misattributed virtual objects as real at rates approaching 20\%~\cite{bonnail2024real}, diverged behaviorally when navigating holographic versus physical obstacles~\cite{coolen2020avoiding}, and expressed concern about their inability to distinguish real hazards from virtual ones~\cite{lebeck2018towards}. These observations align with models of perceptual illusion maintenance through multisensory integration and predictive processing~\cite{gonzalez2017model, westermeier2023exploring, brubach2024manipulating}, and with analyses suggesting that remaining discrimination cues will progressively erode as rendering advances~\cite{kruijff2010perceptual, nijholt2023toward}.

We argue that this growing inability to discriminate virtual from physical content constitutes a \textbf{virtual-physical discrimination vulnerability} that adversaries can systematically exploit. Prior XR security research has examined environment-level manipulations in VR~\cite{casey2019immersive, yang2024inception}, platform UI vulnerabilities~\cite{cheng2024user, mukherjee2025shadowed}, shared-state poisoning~\cite{slocum2024doesn}, multi-sensory perceptual manipulation~\cite{cheng2023exploring}, and XR-specific dark patterns~\cite{hadan2024deceived, krauss2024makes}. However, none systematically investigates how attackers can exploit virtual-physical confusion \textit{at the object level} in MR to cause users to misjudge whether specific objects are real and act on those misjudgments with physical-world consequences. We address this gap by investigating \textbf{virtual-physical confusion attacks}, in which adversaries inject or overlay deceptive virtual content that users mistake for physical reality or misinterpret as genuine attributes of real objects. Two research questions guide our work:

\begin{itemize}
    \item \textbf{RQ1:} What is the design space of attacks exploiting virtual-physical discrimination vulnerability, and how can these attacks be systematically categorized?
    \item \textbf{RQ2:} What are the effects of these attacks on users' perception, cognition, and behavior, and how do users react when subjected to such attacks?
\end{itemize}

We adopt a two-phase approach. To address RQ1, we conducted three speculative design workshops with 12 experts from cybersecurity and MR/HCI, deriving from 36 generated scenarios a taxonomy of two attack classes (\textit{Injection} and \textit{Overlay}) with four subtypes that capture distinct mechanisms of virtual-physical deception. To address RQ2, we implemented four proof-of-concept attacks on Apple Vision Pro using standard platform APIs, each instantiating one subtype within a realistic MR task, and evaluated them with 26 participants through behavioral observation, questionnaires, and semi-structured interviews.

All four attacks altered user behavior at 85--100\% success rates. The most successful attacks were also the hardest to detect, leaving no internal signal to trigger suspicion. Even when participants recognized virtual overlays as artificial, the overlays' influence on perceived object types persisted, and obvious fakes created protective blind spots. Detection did not prevent compliance: 88\% detoured around virtual obstacles they recognized as non-physical, and those who identified product overlays as virtual still used them as decision criteria. No participant attributed anomalous events to adversarial causes. Our contributions are:

\begin{itemize}
    \item We formalize virtual-physical discrimination vulnerability as an exploitable security primitive and present a four-subtype attack taxonomy grounded in expert speculative design workshops.
    \item We implement and empirically evaluate four proof-of-concept attacks on a consumer MR platform, demonstrating feasibility and behavioral impact through a controlled study with 26 participants.
    \item We characterize users' discrimination strategies, identify cognitive mechanisms through which attacks succeed even when detected, and propose countermeasures spanning platform-level provenance, interaction gating, and user education.
\end{itemize}

%-------------------------------------------------------------------------------

\section{Related Work}

\subsection{Perceptual Manipulation Attacks in XR}

Extended reality platforms introduce attack surfaces absent from traditional computing~\cite{abraham2022implications, gugenheimer2022novel}. Prior attacks operate at three levels. At the \textit{environment} level, the Human Joystick attack exploited unprotected VR safety configurations to steer users to attacker-designated locations~\cite{casey2019immersive}, and the Inception attack replicated an entire VR interface as a man-in-the-middle layer~\cite{yang2024inception}. At the \textit{system interface} level, AR/MR platforms permit synthetic input without provenance~\cite{cheng2024user}, shared-state frameworks are vulnerable to remote hologram injection~\cite{slocum2024doesn}, and WebXR ecosystems are susceptible to cursor-jacking~\cite{lee2021adcube, mukherjee2025shadowed}. Defensive approaches span platform output policies~\cite{lebeck2017securing}, per-object access control~\cite{ruth2019secure}, provenance-based auditing~\cite{shoaib2025principled}, and VLM-based manipulation detection~\cite{xiu2025detecting, xiu2025viddar}. Table~\ref{tab:positioning} provides a structured comparison across dimensions relevant to our contribution.

At the \textit{sensory channel} level, the work most closely related to ours, Cheng et al. defined Perceptual Manipulation Attacks (PMA) and demonstrated that visual, auditory, and situational-awareness manipulations induce reaction-time delays and misattribution of anomalies to system malfunctions~\cite{cheng2023exploring}. Tseng et al. proposed a Puppetry/Mismatching taxonomy of Virtual-Physical Perceptual Manipulations through speculative workshops, but without empirical validation on MR platforms~\cite{tseng2022dark}. SwitchAR showed that pass-through AR users can be covertly switched to a photogrammetric reconstruction with zero spontaneous detection~\cite{wombacher2025switchar}. Most recently, the SoK by Teymourian et al. unified MR security, information theory, and cognition into a deception analysis framework, explicitly calling for empirical validation of deception's cognitive effects~\cite{teymourian2025sok}.

Our work addresses a gap that cuts across these contributions. PMA evaluates sensory-channel manipulations through microbenchmark tasks rather than ecologically grounded scenarios; the VPPM taxonomy remains speculative and VR-only; SwitchAR operates at the environment level by replacing the entire reality feed. None targets users' \textit{object-level ontological judgment}: whether a specific object in their surroundings is real or virtual, and whether its perceived type or attributes are genuine. We focus on this object-level confusion in MR, providing a four-subtype attack taxonomy, proof-of-concept implementations on consumer hardware, and empirical measurement of behavioral impacts, directly addressing the gap identified by Teymourian et al.~\cite{teymourian2025sok}.

%------- summary table -------
\begin{table*}[t]
\centering
\caption{Related-work positioning matrix. \textit{Setting}: VR, AR, or MR. \textit{Granularity}: manipulation level (Env.=environment/reference-frame, Sys.=system interface, UI=virtual UI, Obj.=object identity/attributes, Sensory=sensory channel). \textit{V-P Confusion}: whether virtual-physical boundary judgment is a primary attack target (\checkmark), partially relevant ($\triangle$, e.g., confusion observed but not systematically targeted), or not addressed (--). \textit{Taxonomy}: structured attack classification provided. \textit{PoC on MR}: attacks implemented on consumer MR hardware. \textit{User Study}: empirical evaluation under attack conditions with participant count.}
\label{tab:positioning}
\small
\setlength{\tabcolsep}{4pt}
\begin{tabular}{@{}l l c c c c c c@{}}
\toprule
\textbf{Work} & \textbf{Venue} & \textbf{Setting} & \textbf{Granularity} & \textbf{V-P Confusion} & \textbf{Taxonomy} & \textbf{PoC on MR} & \textbf{User Study} \\
\midrule
Lebeck et al.~\cite{lebeck2018towards}          & IEEE S\&P '18        & AR        & Obj.          & $\triangle$   & --            & --            & \checkmark~(n=22) \\
Casey et al.~\cite{casey2019immersive}           & IEEE TDSC '19        & VR        & Env./Sys.     & --            & --            & --            & \checkmark~(n=64) \\
Tseng et al.~\cite{tseng2022dark}                & CHI '22              & VR        & Env./Obj.     & $\triangle$   & \checkmark    & --            & -- \\
Cheng et al.~\cite{cheng2023exploring}           & USENIX Sec. '23      & MR        & Sensory/UI    & $\triangle$   & --            & \checkmark    & \checkmark~(n=21) \\
Wang et al.~\cite{wang2024dark}                  & IJHCI '23            & AR        & UI/Obj.       & $\triangle$   & $\triangle$   & --            & \checkmark~(n=15) \\
Cheng et al.~\cite{cheng2024user}                & USENIX Sec. '24      & AR        & UI/Sys.       & --            & $\triangle$   & $\triangle$   & -- \\
Slocum et al.~\cite{slocum2024doesn}             & USENIX Sec. '24      & AR        & Sys./Obj.     & $\triangle$   & $\triangle$   & $\triangle$   & -- \\
Hadan et al.~\cite{hadan2024deceived}            & ACM CSUR '24         & XR        & --            & $\triangle$   & \checkmark    & --            & -- \\
Mukherjee et al.~\cite{mukherjee2025shadowed}    & USENIX Sec. '25      & WebXR & UI/Sys.       & --            & \checkmark    & --            & \checkmark~(n=100) \\
Wombacher et al.~\cite{wombacher2025switchar}    & UIST '25             & AR     & Env.          & \checkmark    & --            & \checkmark    & \checkmark~(n=20) \\
Teymourian et al.~\cite{teymourian2025sok}       & USENIX Sec. '25      & MR        & --            & $\triangle$   & \checkmark    & --            & -- \\
Sajid et al.~\cite{sajid2025just}                & IEEE VR '25          & MR        & UI/Sys.       & --            & $\triangle$   & \checkmark    & \checkmark~(n=20) \\
\midrule
\textbf{Our work}                                &                    & MR        & Obj.          & \checkmark    & \checkmark    & \checkmark    & \checkmark~(n=26) \\
\bottomrule
\end{tabular}
\vspace{-3mm}
\end{table*}

\subsection{Dark Patterns and Deceptive Design in XR}
Systematic reviews and expert co-design studies have established that XR introduces qualitatively new manipulation vectors. Hadan et al. identified 15 subthemes of deceptive design including reality distortion and perception tricking~\cite{hadan2024deceived}, while Krauss et al. produced 42 dark pattern scenarios and specifically noted that indistinguishable AR content could lead users to believe virtual objects physically exist~\cite{krauss2024makes}. 
Further work covers AR retail~\cite{ruocco2024redirected}, advertising leveraging realism~\cite{mhaidli2021identifying, mhaidli2023shockvertising}, co-located credibility effects~\cite{wang2024dark, eghtebas2023co}, dark pattern prevalence in 80 MR apps~\cite{todhriimmersion}, and XR memory manipulation~\cite{bonnail2023memory}.
However, these contributions predominantly remain at the speculative, survey, or scenario-construction level. We operationalize the identified risks as adversarial attacks with a formal threat model, implement them on consumer MR hardware, and empirically measure both behavioral outcomes and user defensive responses.

\subsection{Virtual-Physical Perceptual Ambiguity}
Users confuse virtual content with physical reality even without adversarial intent: 20\% of VR participants sat on virtual chairs without verification~\cite{wiesing2025confusing}, source judgment tests show $\sim$20\% virtual-to-real misattribution~\cite{bonnail2024real}, MR users navigate holographic obstacles differently than real ones~\cite{coolen2020avoiding}, and AR users routinely treat holograms as physical entities~\cite{lebeck2018towards}. Recent work explores visual cues to reduce source confusion~\cite{petiot2025using}.

These perceptual failures are well-grounded in models of multisensory integration, top-down mismatch suppression, and plausibility judgment~\cite{gonzalez2017model, westermeier2023exploring, brubach2024manipulating}, and are expected to worsen as the AR community pursues indistinguishability through advances in rendering~\cite{itoh2021towards, pardo2018correlation} and deep-learning-based object insertion~\cite{niu2021making, kruijff2010perceptual, nijholt2023toward}. Prior work treats this ambiguity as a usability challenge or cognitive phenomenon. We reframe it as a security vulnerability that adversaries can systematically exploit, and provide the first empirical characterization of how virtual-physical confusion attacks affect user behavior in realistic MR tasks.

% ============================================================
% STUDY 1: SPECULATIVE DESIGN WORKSHOPS — METHOD
% ============================================================

% (\textit{How might adversaries exploit users' vulnerability in discriminating virtual from physical content to mount attacks?})

\section{Study 1: Speculative Design Workshops}

To answer RQ1, we conducted speculative design workshops with domain experts. Speculative design has proven effective at surfacing threat scenarios in XR security research, where attacks depend on the interplay between human perception and platform affordances~\cite{tseng2022dark, bonnail2023memory, krauss2024makes, ruocco2024redirected, eghtebas2023co}. % We adapted this methodology to the domain of virtual-physical confusion in MR.

\subsection{Method}

\subsubsection{Participants}

We recruited 12 researchers (3 women, 9 men) through purposive sampling from two complementary domains: MR/HCI and cybersecurity. Nine had expertise in MR or HCI, seven in cybersecurity, and four held cross-domain expertise spanning both areas. This composition follows prior XR threat elicitation studies~\cite{tseng2022dark, bonnail2023memory, krauss2024makes} and ensures that generated scenarios are grounded in both MR perceptual affordances and realistic adversarial reasoning. All participants were active researchers with publications at top-tier venues in their respective fields (see Table~\ref{tab:workshop_participants}). Self-rated MR familiarity on a five-point scale ranged from average (n=3) to above average (n=5) to expert (n=4). Each participant received \$50 compensation.

\subsubsection{Procedure}
We conducted three workshops of four participants each (Table~\ref{tab:workshop_participants}), lasting 90--120 minutes via video conference. The study protocol was approved by our university's Institutional Review Board. Each workshop comprised three phases: pre-workshop preparation, brainstorming, and group discussion.

Each participant received background materials 3--4 days before each session and independently designed three attack scenarios specifying target, context, procedure, and harms~\cite{teymourian2025sok, hadan2024deceived, bonnail2023memory, ruocco2024redirected, mhaidli2021identifying}. Sessions comprised a brainstorming phase ($\sim$60\,min) and a collective discussion across MR-specific capabilities, preventive measures, likelihood, and attacker incentives ($\sim$45\,min). All sessions were audio-recorded and transcribed with consent. Full materials are in Appendix~\ref{sec:workshop_protocol}.

\subsubsection{Data Analysis}
We analyzed workshop transcripts using hybrid thematic analysis~\cite{swain2018hybrid} with incremental open coding~\cite{corbin1990grounded}. Two researchers independently coded all 36 scenarios along three dimensions: the relationship between virtual content and physical objects, the mechanism of deception, and the intended effect on user perception and behavior (Cohen's $\kappa = 0.81$, substantial agreement~\cite{landis1977measurement}). This analysis produced the attack taxonomy in Section~\ref{sec:taxonomy}, in which categories emerged from the data rather than from a predetermined framework.

\begin{table}[t]
\centering
% \vspace{-5mm}
\caption{Workshop participant expertise. MR familiarity: $\bullet\!\bullet\!\bullet$ = Expert, $\bullet\!\bullet$ = Above average, $\bullet$ = Average.}
\label{tab:workshop_participants}
\footnotesize
\setlength{\tabcolsep}{2.5pt}
\begin{tabularx}{\columnwidth}{@{}lclcX@{}}
\toprule
\textbf{ID} & \textbf{G} & \textbf{Expertise} & \textbf{MR Fam.} & \textbf{Publication Venues} \\
\midrule
W1P1 & M & Security, MR/HCI & $\bullet\!\bullet\!\bullet$ & CHI, ISMAR, TVCG \\
W1P2 & M & Security, HCI  & $\bullet$ & CCS, CHI, Ubicomp, CSCW \\
W1P3 & M & MR/HCI & $\bullet\!\bullet$ & CHI \\
W1P4 & F & XR/HCI, Design & $\bullet\!\bullet$ & SIGGRAPH Asia, CHI \\
\midrule
W2P5 & M & Security, MR/HCI & $\bullet\!\bullet$ & TVCG, ACCV \\
W2P6 & F & Security, HCI  & $\bullet$ & IEEE S\&P, NDSS, ToCHI \\
W2P7 & F & MR/HCI, Design & $\bullet\!\bullet\!\bullet$ & IEEE VR, VRST \\
W2P8 & M & MR/HCI, Security & $\bullet\!\bullet$ & UIST, CHI, ISMAR \\
\midrule
W3P9 & M & Security, MR & $\bullet\!\bullet\!\bullet$ & TVCG, IEEE VR, ISMAR \\
W3P10 & M & MR & $\bullet\!\bullet\!\bullet$ & Ubicomp, ISMAR \\
W3P11 & M & MR/HCI & $\bullet\!\bullet\!\bullet$ & IEEE Internet Comput. \\
W3P12 & M & Security, AI & $\bullet$ & CCS, ICLR, ECCV \\
\bottomrule
\end{tabularx}
\vspace{-5mm}
\end{table}

\subsection{Results}

\begin{table*}[t]
\centering
\caption{Taxonomy of virtual-physical confusion attacks with formal definitions and representative workshop scenarios. }
\label{tab:taxonomy}
\small
\begin{tabular}{p{2.0cm} p{2.2cm} p{4.8cm} p{6cm}}
\toprule
\textbf{Class} & \textbf{Subtype} & \textbf{Formal Definition} & \textbf{Representative Scenarios} \\
\midrule
\multirow{6}{2.0cm}{Injection\\($O_r \subsetneq O_m$)}
& Endogenous & Cloned duplicates of existing physical objects are injected & Cloned LEGO piece triggers forced ad click; duplicate apples obstruct object retrieval; virtual food on a laptop induces misplacement \\
\cmidrule(l){2-4}
& Exogenous & Pre-built virtual assets with no physical counterpart are injected & Fake signs misdirect navigation; virtual obstacles alter walking paths; virtual pet lures children; fake ``Closed'' sign diverts customers \\
\midrule
\multirow{6}{2.0cm}{Overlay\\($p = q,\; \forall o_m^i: \pi(o_m^i) \neq \emptyset$)}
& Type-Deceptive & Virtual overlay changes the perceived type of a physical object ($\exists\, i: \mathit{type}(o_m^i) \neq \mathit{type}(\pi(o_m^i))$) & Trash can rendered as a flower pot obstructs cleanup; water cup disguised as a pen holder; utensil misidentification during cooking \\
\cmidrule(l){2-4}
& Attribute-Deceptive & Virtual overlay alters perceived attributes without changing recognized type ($\mathit{type}$ preserved, $\mathit{attr}$ distorted) & Fake stains reduce product appeal; brand logos inflate perceived value; misleading price labels; altered team colors cause friendly-fire confusion \\
\bottomrule
\end{tabular}
\end{table*}

\begin{figure*}[t]
\centering
\includegraphics[width=\linewidth]{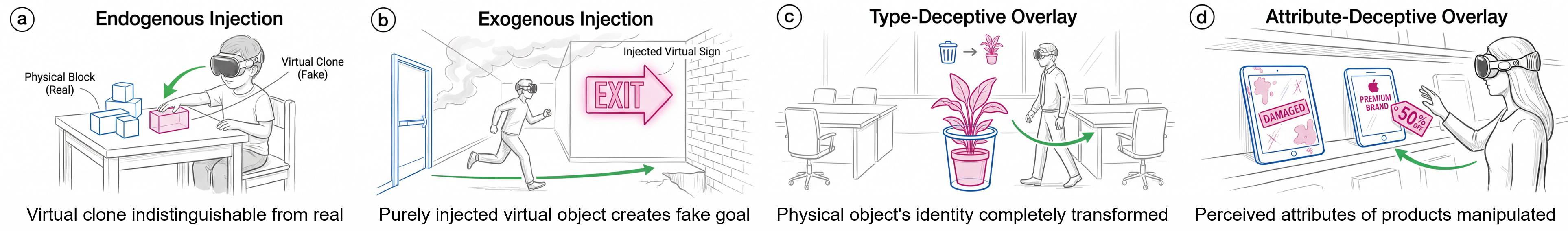}
\vspace{-7mm}
\caption{Illustrative scenarios for the four subtypes of virtual-physical confusion attacks. Blue outlines represent physical objects; pink/magenta indicates virtual content rendered by the MR headset; green arrows show users' misled actions.}
\label{fig:taxonomy_illustration}
\vspace{-3mm}
\end{figure*}

\subsubsection{Scenario Overview}
The 36 scenarios (Appendix~\ref{appendix:scenarios}) spanned diverse contexts (navigation, shopping, collaboration, assembly, evacuation, healthcare, gaming, social interaction), target populations (consumers, office workers, children, surgical teams, drivers), and harms (physical injury, financial loss, task failure, privacy violation, psychological distress).

\subsubsection{Attack Taxonomy}
\label{sec:taxonomy}

Thematic analysis produced a two-class, four-subtype taxonomy of \textit{virtual-physical confusion attacks} (Table~\ref{tab:taxonomy}). Let $O_r = \{o_r^1, \ldots, o_r^p\}$ be the physical objects and $O_m = \{o_m^1, \ldots, o_m^q\}$ the MR-mediated objects. Each object has a type $\mathit{type}(o)$ (e.g., cup) and attributes $\mathit{attr}(o)$ (color, brand, surface condition). A grounding function $\pi: O_m \rightarrow O_r \cup \{\emptyset\}$ maps each MR-perceived object to its physical counterpart, or to $\emptyset$ if purely virtual.
% A worked example illustrating these definitions applied to our four PoC attacks appears in Appendix~\ref{appendix:worked_example}.

\textbf{Injection attacks}
introduce new virtual objects into the MR-perceived environment while leaving physical objects unaltered. Formally, $O_r \subsetneq O_m$: one or more purely virtual objects have been added ($\exists\, o_m^i \in O_m$ s.t. $\pi(o_m^i) = \emptyset$). We distinguish two subtypes by the origin of the injected content.

\textit{Endogenous injection} attacks clone objects already present in the physical scene and re-introduce them as virtual duplicates sharing the type and attributes of their physical counterparts. Because duplicates are visually indistinguishable from the originals, users cannot determine which instance is real. In a representative scenario (Figure~\ref{fig:taxonomy_illustration}a), a LEGO assembly assistant clones a physical block and renders a virtual duplicate elsewhere; grasping the clone triggers a hidden ad click (W1P1).

\textit{Exogenous injection} attacks introduce pre-constructed virtual assets with no physical counterpart, altering the environmental semantics. In a representative scenario (Figure~\ref{fig:taxonomy_illustration}b), fake directional signs misdirect pedestrians toward dangerous areas (W1P3, W3P12); other instances include virtual obstacles (W3P9), fake emergency exits (W1P4), and a virtual ``Closed'' sign diverting customers (W3P11). Scenarios in this subtype clustered around directional manipulation; other exogenous forms likely exist but did not surface in our workshops. These attacks succeed because users lack prior expectations about which objects should be present, making injected content difficult to question.

\textbf{Overlay attacks}
superimpose virtual content onto existing physical objects to alter how those objects are perceived, without introducing new objects. The object count remains unchanged ($p = q$), and every MR-perceived object has a physical counterpart ($\forall\, o_m^i,\; \pi(o_m^i) \neq \emptyset$), but the overlay creates a discrepancy between perceived and actual properties. We distinguish two subtypes by whether the overlay changes perceived object type or only its attributes.

\textit{Type-deceptive overlay} attacks alter a physical object's appearance so substantially that the user perceives it as a different type of object: $\exists\, i:\; \mathit{type}(\pi(o_m^i)) \neq \mathit{type}(o_m^i)$. In a representative scenario (Figure~\ref{fig:taxonomy_illustration}c), a MR assistant renders virtual flowers over a trash can, causing users to perceive it as a flower pot and abandon the associated cleanup subtask (W1P1). These attacks exploit the dominance of visual appearance in MR object recognition: a sufficiently convincing overlay overrides contextual and spatial cues.

\textit{Attribute-deceptive overlay} attacks modify the perceived attributes of a physical object without changing its recognized type: $\mathit{type}(\pi(o_m^i)) = \mathit{type}(o_m^i)$, but $\mathit{attr}(\pi(o_m^i)) \neq \mathit{attr}(o_m^i)$. The user correctly identifies what the object is but forms an incorrect judgment about its properties. This subtype was most commonly proposed in shopping contexts (Figure~\ref{fig:taxonomy_illustration}d): virtual stains or scratches reduced product appeal, while brand logos inflated perceived value (W1P2) and misleading price labels created false urgency (W2P8, W3P12). Unlike type-deceptive overlays, attribute-deceptive overlays are subtler: basic object identification remains correct, but the attack distorts evaluative judgments that inform downstream decisions.

\section{Proof-of-Concept Attacks}
\label{poc_attacks}

To empirically validate the attack taxonomy from Study~1, we designed four proof-of-concept (PoC) attacks, each instantiating one taxonomy subtype within a distinct MR task. All four were implemented on Apple Vision Pro running visionOS 26, using ARKit (world tracking, plane tracking, hand tracking), RealityKit (3D rendering and material processing), and SwiftUI.

\begin{figure}[t]
\centering
\includegraphics[width=0.9\columnwidth]{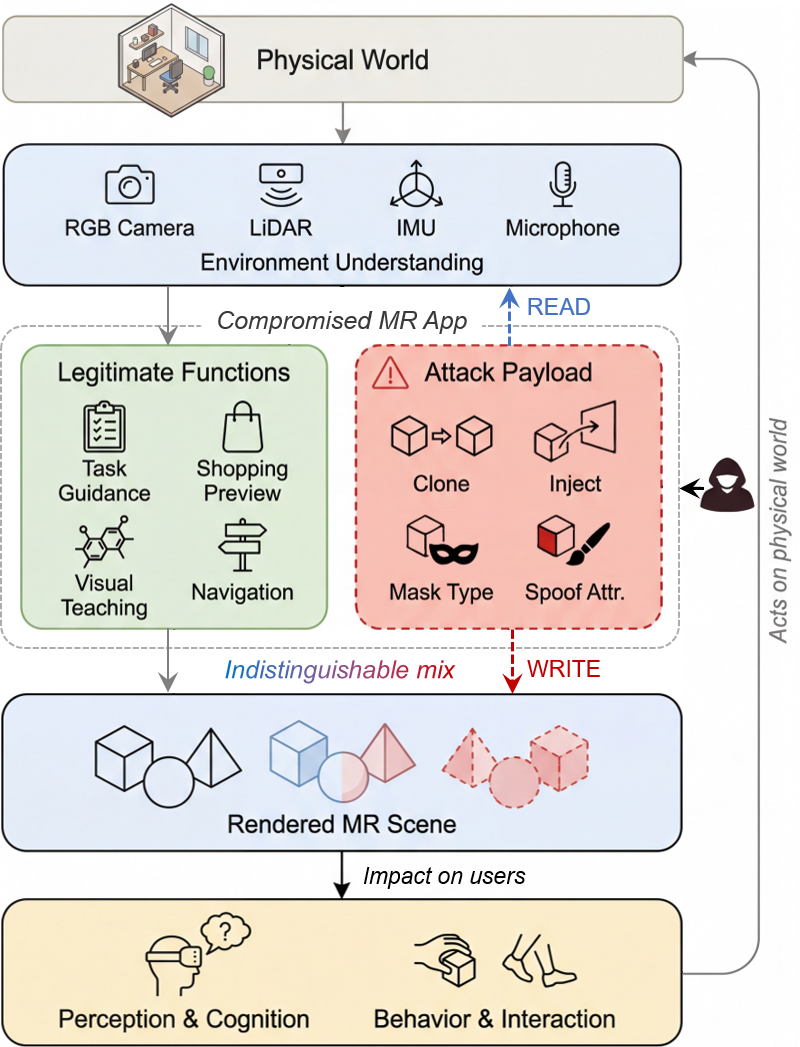}
\vspace{-2mm}
\caption{Threat model overview. A compromised MR task-assistance app retains its legitimate functions (green) while harboring an attack payload (red). The attacker \textcolor{blue}{\textbf{reads}} environment data from the device perception layer and \textcolor{red}{\textbf{writes}} deceptive virtual content into the rendering pipeline via four attack subtypes. Legitimate and malicious content merge into an indistinguishable MR scene, affecting user perception, cognition, and physical-world behavior.}
\vspace{-3mm}
\label{fig:threat_model}
\end{figure}

\subsection{Threat Model and Attack Preconditions}
\label{sec:threat_model}

\textbf{Attacker foothold.}
The attacker operates through a compromised MR task-assistance application (Figure~\ref{fig:threat_model}). This is a realistic foothold given documented vectors: malicious SDKs embedded in legitimate applications~\cite{cheng2024user, mukherjee2025shadowed}, supply-chain compromise of app backends, and platform vulnerabilities permitting unauthorized content injection~\cite{asif2024breaking}. Because the application retains its legitimate permissions and user trust, users interact with it expecting helpful task guidance while the rendering pipeline has been subverted. Attacker motivations span the four scenarios we evaluate: ad revenue (BuildAssist), biased purchasing (ShopLens), task disruption (TidySpace), and misdirection of physical movement (PathGuide).

\textbf{Attacker capabilities.}
The compromised application possesses two capabilities (Figure~\ref{fig:threat_model}). First, it can \textit{read} world-sensing outputs from the device's perception pipeline, including spatial maps, plane geometry, and tracked object positions. On Apple Vision Pro, applications granted ARKit access receive spatially registered environment data within a limited radius~\cite{apple2025objecttracking}. Second, it can \textit{write} arbitrary virtual content into the MR scene with full control over geometry, texture, spatial anchoring, and occlusion, and can register interaction callbacks so that user proximity or touch triggers application logic~\cite{cheng2024user, sajid2025just}. None of the four attacks requires raw camera frame access; all operate through standard rendering APIs, consistent with the principle that AR output manipulation is achievable without privileged sensor access~\cite{lebeck2017securing}.

As a concrete example, a third-party spatial analytics SDK integrated into an MR app could legitimately request ARKit and Full Space permissions for usage heatmaps while containing an obfuscated payload that reads object positions and injects deceptive content at runtime, requiring no additional privileges. Analogous SDK-based supply-chain attacks have been documented in mobile~\cite{cheng2024user} and WebXR ecosystems~\cite{mukherjee2025shadowed}.

\textbf{Preconditions.}
Two preconditions enable the attacks: (i) the MR platform renders virtual content with sufficient fidelity that users cannot reliably distinguish it from physical objects --- met by current consumer devices for spatially anchored, texture-matched content; (ii) virtual content coexists with the physical environment, a condition amplified when multiple applications share the same spatial scene~\cite{zhou2024spatial}.

\subsection{Disguised Ad Attack on \textit{BuildAssist}}
\label{sec:attack_a}

\textit{BuildAssist} is an MR assembly assistant that renders a high-fidelity 3D reference model of the target structure and provides real-time guidance for physical block-building tasks.

\begin{figure}[t]
\centering
\includegraphics[width=\columnwidth]{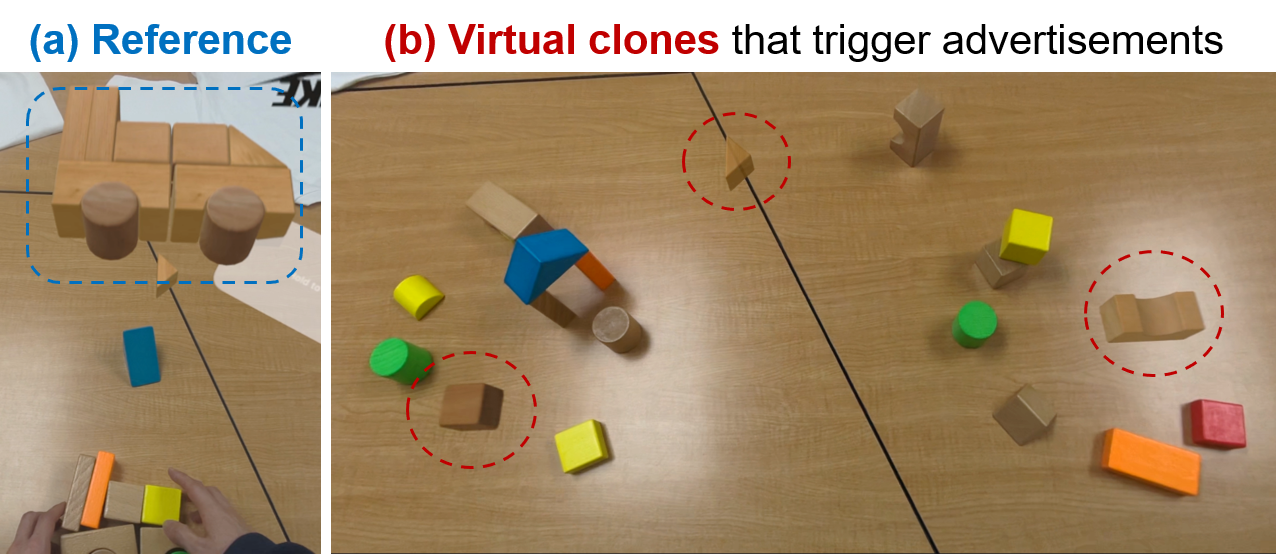}
\vspace{-6mm}
\caption{Disguised Ad attack on \textit{BuildAssist}. (a) The 3D reference model guides block assembly. (b) Three virtual clones of physical blocks (red dashed circles) are injected among real blocks. Grasping a clone triggers a video advertisement.}
\label{fig:attack_a}
\vspace{-3mm}
\end{figure}

\textbf{Attack design (Endogenous Injection).}
The attacker leverages environmental sensing to identify the positions and types of physical blocks, then injects three virtual clones replicating existing blocks in geometry, color, and surface texture (Figure~\ref{fig:attack_a}). Clones are spatially registered to the table surface with correct occlusion and shadow behavior. Each clone functions as a disguised interaction trigger: when the user's hand collides with a clone while attempting to grasp what they believe is a physical block, the system registers a tap event and launches a 10-second video advertisement anchored 1\,m in front of the user. The clone disappears upon contact. This attack extends the cursor-jacking paradigm~\cite{lee2021adcube} from virtual UI elements to cloned physical objects. An attack instance is logged as successful when the system records a hand-collision event with any virtual clone.

\begin{figure}[t]
\centering
\includegraphics[width=\columnwidth]{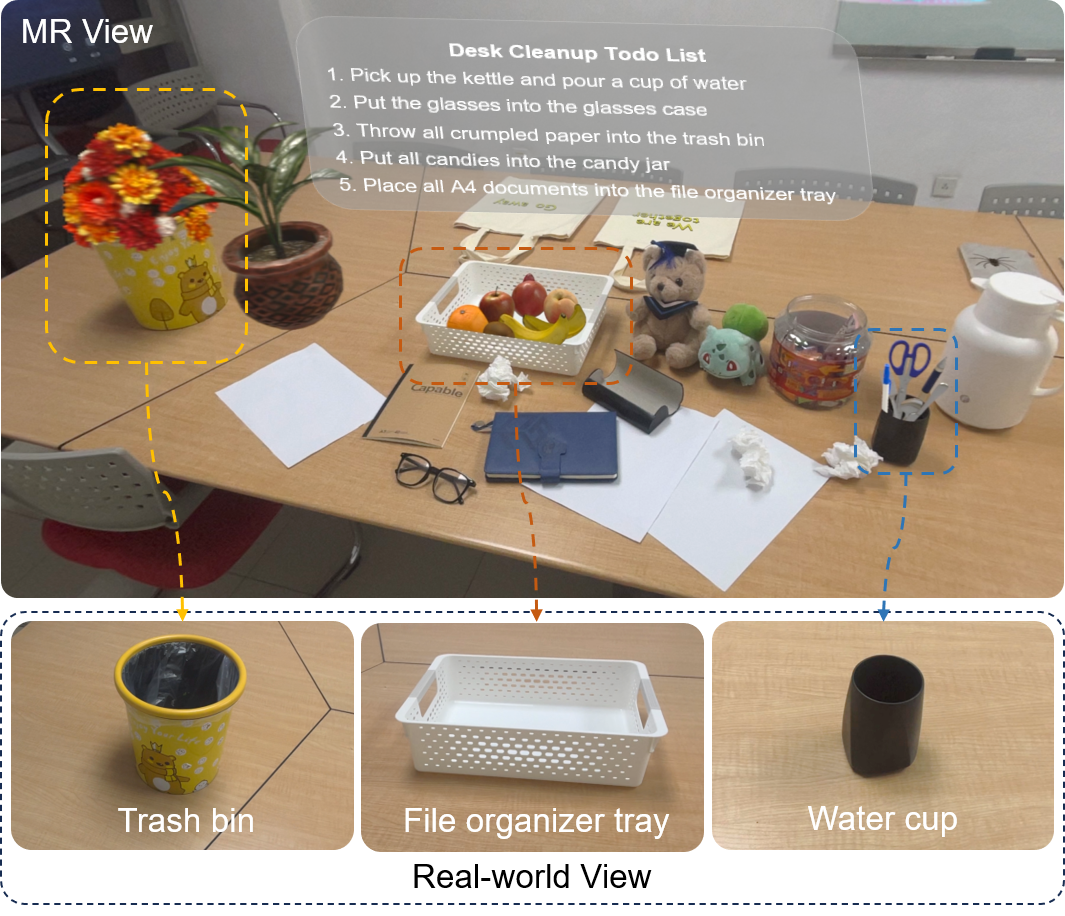}
\vspace{-5mm}
\caption{Object Masquerade attack on \textit{TidySpace}. Top: the user's MR view showing a to-do list and the manipulated environment. Virtual overlays disguise a trash bin as a flower pot (orange), a file organizer tray as a fruit basket (orange), and a water cup as a pen holder (blue). Bottom: the three target objects shown without overlays.}
\label{fig:attack_b}
\vspace{-3mm}
\end{figure}

\begin{figure}[t]
\centering
\includegraphics[width=\columnwidth]{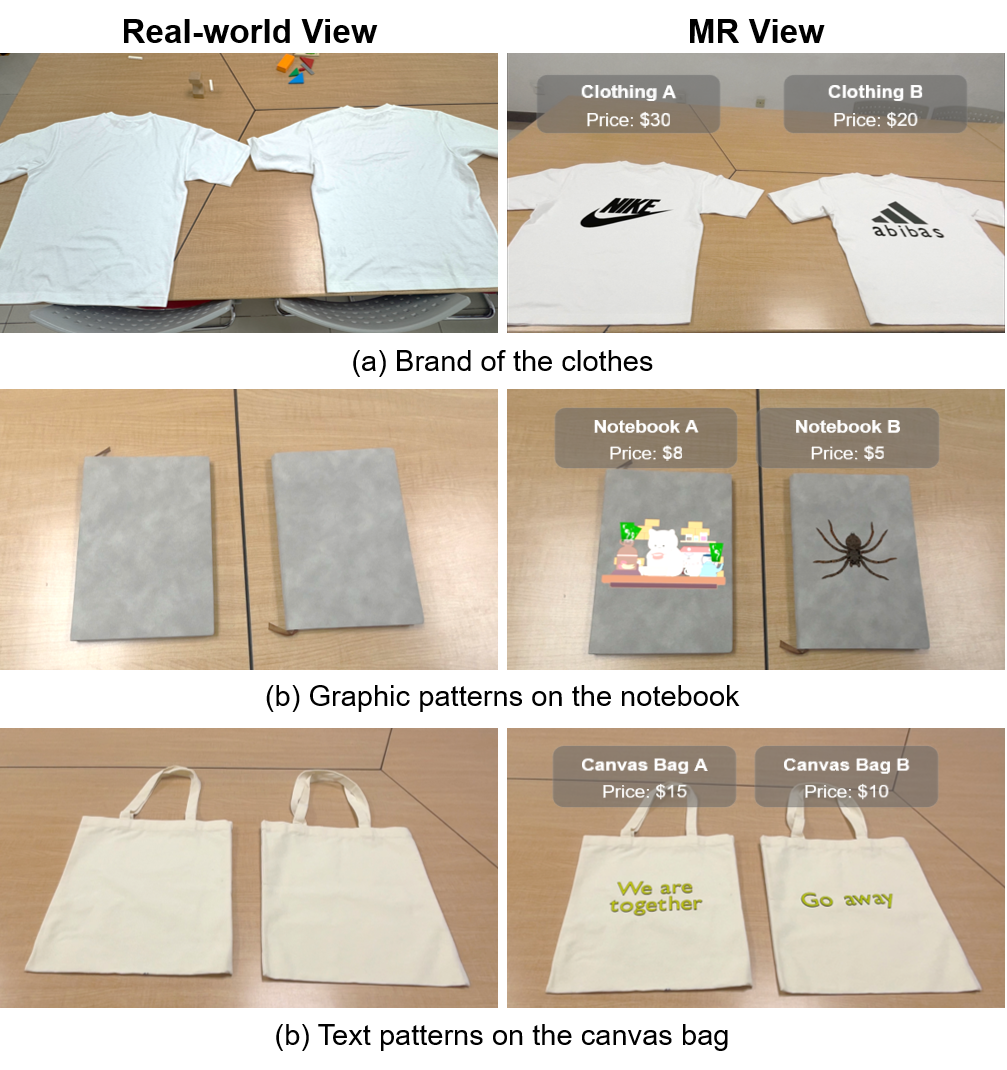}
\vspace{-7mm}
\caption{Surface Spoof attack on \textit{ShopLens}. Left column: real-world view of physically identical product pairs. Right column: MR view with attribute-deceptive overlays. (a) Brand logos on T-shirts. (b) Graphic patterns on notebooks. (c) Text slogans on canvas bags. Higher-priced items receive attractive overlays; lower-priced items receive aversive ones.}
\label{fig:attack_c}
\vspace{-3mm}
\end{figure}

\subsection{Object Masquerade Attack on \textit{TidySpace}}
\label{sec:attack_b}

\textit{TidySpace} is an MR desk cleanup assistant that displays a floating to-do list of sequential organizing subtasks (e.g., pour water into a cup, discard crumpled paper into the trash bin, file documents into the organizer tray).

\textbf{Attack design (Type-Deceptive Overlay).}
The attacker overlays high-fidelity virtual elements onto three physical objects to alter their perceived functional identity (Figure~\ref{fig:attack_b}): virtual flowers on the trash bin (appearing as a flower pot), virtual fruit in the file organizer tray (appearing as a fruit basket), and virtual pens and scissors in the water cup (appearing as a pen holder). Each overlay preserves the spatial position and physical boundaries of the underlying object while rewriting its perceived category, preventing task completion by making target containers unrecognizable. An attack instance is logged as successful when a participant abandons a subtask, verbally concluding that the required object is not present.

\subsection{Surface Spoof Attack on \textit{ShopLens}}
\label{sec:attack_c}

\textit{ShopLens} is an MR shopping assistant that overlays digital information panels (product name, price) above physical merchandise to support purchasing decisions.

\textbf{Attack design (Attribute-Deceptive Overlay).}
The attacker overlays deceptive visual attributes onto physically identical product pairs to bias purchasing decisions toward higher-priced items (Figure~\ref{fig:attack_c}). Across three product categories, the higher-priced option receives an attractiveness-enhancing overlay while the lower-priced option receives a repulsive one: (1)~T-shirts with a recognizable brand logo (``Nike'') versus a counterfeit-brand logo (``Abibas''); (2)~notebooks with an appealing illustration versus an aversive graphic pattern; (3)~canvas bags with a positive slogan versus a hostile message. The underlying physical products in each pair are identical, differing only in virtually rendered surface attributes. After each selection, participants verbally explain their choice, enabling analysis of whether rationales reference virtual attributes. An attack instance is logged as successful when a participant's selection or reasoning is demonstrably driven by the manipulated overlays.

\subsection{Phantom Obstacle Attack on \textit{PathGuide}}
\label{sec:attack_d}

\textit{PathGuide} is an MR indoor navigation assistant that renders dynamic directional arrows on the floor and animated destination markers to guide users between waypoints.

\begin{figure}[t]
\centering
\includegraphics[width=\columnwidth]{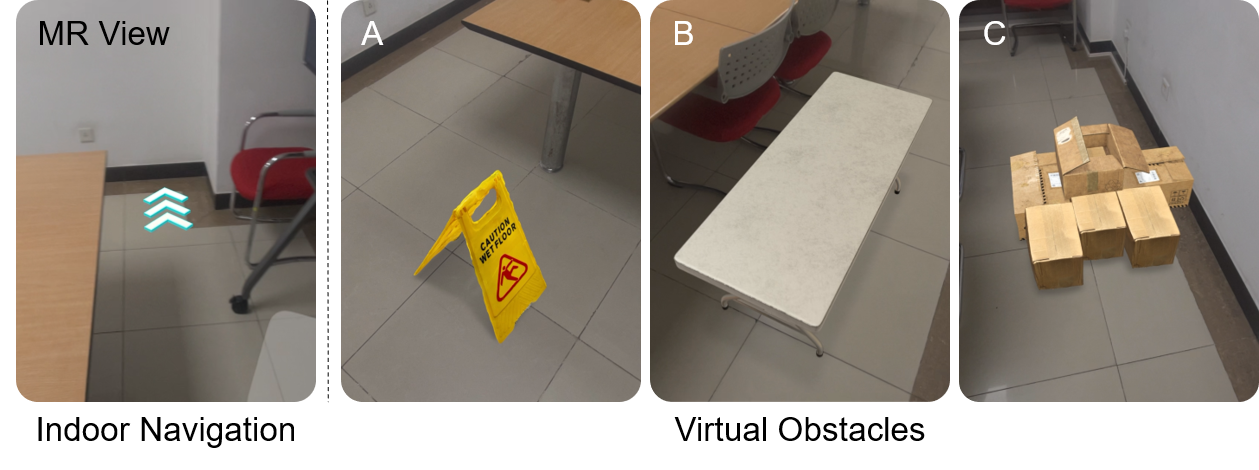}
\vspace{-7mm}
\caption{Phantom Obstacle attack on \textit{PathGuide}. Left: the user's MR view showing floor-projected navigation arrows. A-C: three virtual obstacles (Wet-floor caution sign, Folding table, and Stack of shipping boxes) injected along the path, rendered with environment-consistent lighting and occlusion.}
\label{fig:attack_d}
\vspace{-3mm}
\end{figure}

\textbf{Attack design (Exogenous Injection).}
The attacker injects three high-fidelity virtual obstacles from a pre-built asset library at strategic points along the user's designated path (Figure~\ref{fig:attack_d}): a wet-floor caution sign, a folding table, and a stack of shipping boxes. Each obstacle is rendered with environment-consistent lighting, shadow casting, and correct occlusion against the physical floor and walls. The attack exploits automatic obstacle-avoidance behavior~\cite{coolen2020avoiding}: users tend to treat plausible obstructions as real hazards and modify their trajectory without consciously evaluating physical presence~\cite{lebeck2018towards, casey2019immersive}. An attack instance is logged as successful when a participant deviates from the designated path to circumvent a virtual obstacle.

\section{Study 2: Evaluating Attack Effectiveness}

To evaluate the feasibility and impact of the four PoC attacks described in Section~\ref{poc_attacks}, we conducted a controlled user study addressing RQ2. The study employed a deception protocol~\cite{cheng2023exploring}: participants were told they were evaluating MR task-assistance applications, and the true nature of the embedded attacks was revealed only during debriefing.

\subsection{Method}
\label{sec:study_2_method}

\textbf{Participants.}
We recruited 26 participants (16 women, 10 men; ages 20--38, $M$=26.33, $SD$=1.53) from a university mailing list. Eligibility required no history of severe motion sickness or photosensitivity. Participants self-rated their XR familiarity on a 5-point scale from 1 (not at all familiar) to 5 (very familiar) ($M$=2.65, $SD$=1.02); the prior-use distribution is reported in Appendix~\ref{appendix:protocol}. Each participant received \$20 for approximately one hour of participation.

\begin{figure}[t]
\centering
\includegraphics[width=\linewidth]{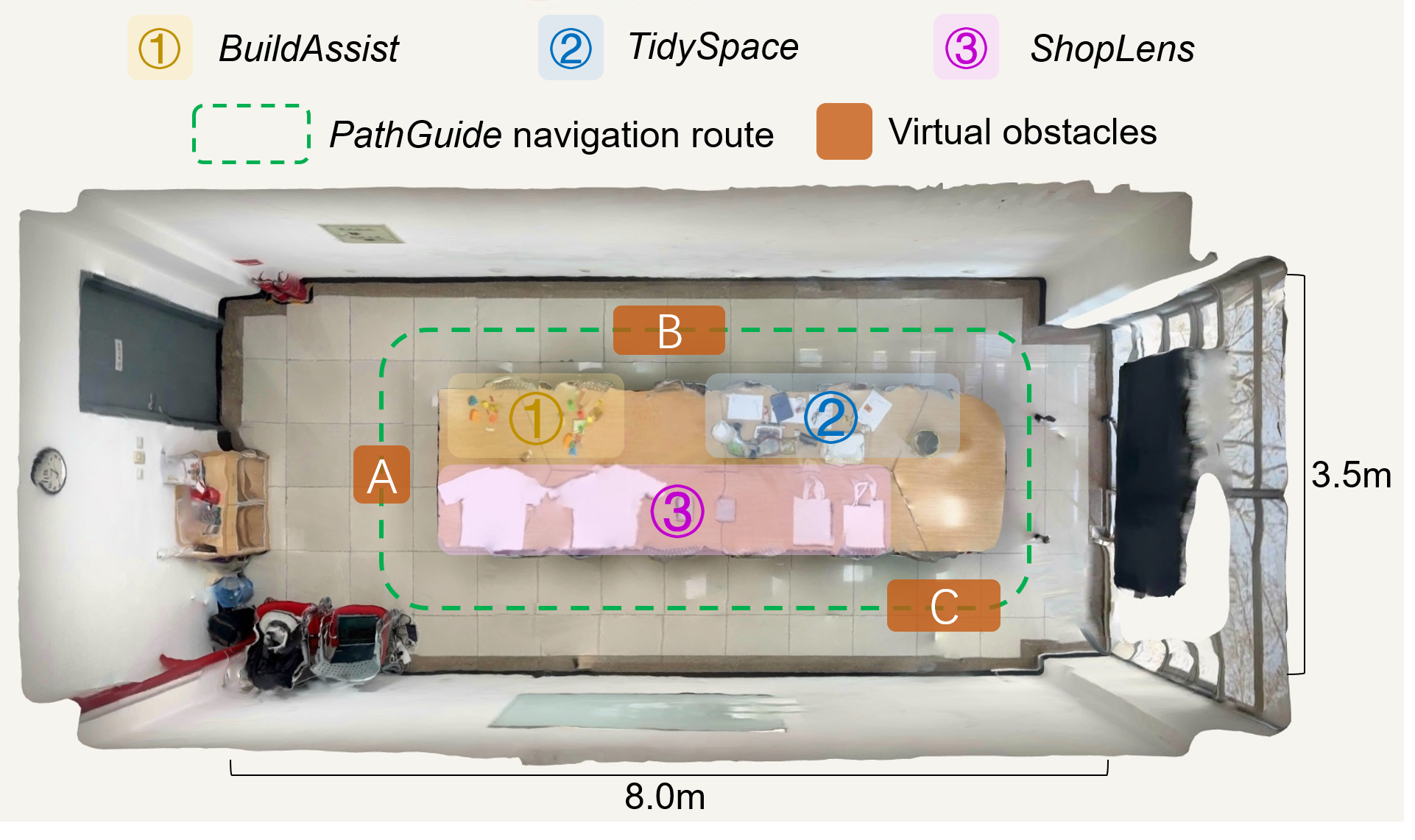}
\vspace{-6mm}
\caption{Top-down view of the experiment space (8.0\,m $\times$ 3.5\,m). Three stationary task stations are labeled: \raisebox{.5pt}{\textcircled{\raisebox{-.9pt} {1}}} block building with \textit{BuildAssist}, \raisebox{.5pt}{\textcircled{\raisebox{-.9pt} {2}}} desk cleanup with \textit{TidySpace}, and \raisebox{.5pt}{\textcircled{\raisebox{-.9pt} {3}}} goods selection with \textit{ShopLens}. The green dashed path shows the navigation route connecting stations, along which virtual obstacles are injected at stationary positions.}
\label{fig:experiment_layout}
\vspace{-3mm}
\end{figure}

\textbf{Design.}
We used a within-subjects design in which every participant completed all four task-attack scenarios. The sequence of the three stationary tasks (block building, desk cleanup, goods selection) was counterbalanced using a Latin square. The navigation task was embedded as the transition between stationary stations, allowing obstacle encounters to occur naturally during walking segments (Figure~\ref{fig:experiment_layout}).

We did not include a no-attack control because behavioral metrics have unambiguous implicit baselines (0\% abandonment on visible targets, $\sim$50\% choice between identical products, 0\% detour on clear paths); adding controls would double session duration and risk fatigue/motion-sickness confounds. This design follows prior XR security studies on novel attack concepts~\cite{cheng2023exploring, wombacher2025switchar, sajid2025just}.

\textbf{Procedure.}
The study comprised four phases (detailed protocol in Appendix~\ref{appendix:study_2}). In the \textit{pre-experiment} phase, participants provided demographics, signed informed consent, and received a cover story framing the study as an evaluation of MR task-assistance technology. During \textit{familiarization} ($\sim$10\,min), participants were fitted with an Apple Vision Pro headset and completed a demo application in an attack-free environment. In the \textit{experimental phase} ($\sim$25\,min), participants completed the three stationary tasks in counterbalanced order, navigating between stations via \textit{PathGuide}. At each station, the MR application provided legitimate task guidance while simultaneously executing its embedded attack. Participants were encouraged to think aloud, and the system continuously logged participant position, hand interactions, voice, and screen-captured the MR view for behavioral coding. In the \textit{post-experiment} phase ($\sim$25\,min), participants completed six validated scales per scenario --- PPQ~\cite{brubach2024manipulating} (plausibility), IPQ~\cite{schubert2001experience} (presence), NASA-TLX~\cite{hart2006nasa} (workload), SUS~\cite{brooke2013sus} (usability), MRC~\cite{katins2024assessing} (security concern), DAF~\cite{teymourian2025sok} (deception impact) --- administered in a matrix layout across scenarios (battery $\sim$12.5\,min/participant), followed by a semi-structured interview ($\sim$20\,min) and a full debriefing.

\textbf{Data collection and analysis.}
For behavioral data, two researchers independently reviewed video recordings of all sessions, coding observable confusion indicators (hesitation before grasping, hand-waving to test object solidity, repeated visual inspection, verbal doubt); disagreements were resolved through discussion. For questionnaire data, we employed Aligned Rank Transform (ART) ANOVA~\cite{elkin2021aligned} for non-parametric factorial analysis with Greenhouse--Geisser corrections where sphericity was violated, and Bonferroni-corrected paired $t$-tests for post-hoc comparisons. Interview recordings were transcribed and analyzed using reflexive thematic analysis~\cite{mcdonald2019reliability, braun2006using}. Complete behavioral coding and interview codebooks are provided in Appendix~\ref{appendix:codebooks}.

\subsection{Ethics Statement}

This study was approved by our university's Institutional Review Board. Because the study employed deception, several safeguards were implemented. All participants provided written informed consent acknowledging audio/video recording and spatial data collection, were informed that MR headset use may cause discomfort, and could withdraw at any time without loss of compensation. Physical safety was maintained throughout: the navigation area was cleared of real hazards, and an experimenter accompanied each participant during walking segments. No participant discontinued the study or reported discomfort. We concluded each session with a structured debriefing revealing the research objectives and methods, and emphasizing that task difficulties resulted from the attack design rather than participant ability. 

We also considered the responsible disclosure implications of publishing proof-of-concept attack designs. The four attacks exploit fundamental perceptual limitations rather than specific software vulnerabilities, so disclosure to a single vendor would not address the underlying issue. We believe that publishing the attack taxonomy and empirical findings serves the broader security community by enabling platform developers to build informed defenses, following the precedent of prior XR security research~\cite{casey2019immersive, cheng2023exploring, wombacher2025switchar}.

\subsection{Results}

\subsubsection{Attack Effectiveness and Behavioral Impact}
\label{sec:effectiveness}

All four attacks succeeded in altering participant behavior, though they differed in mechanism and detectability.

\begin{figure}[t]
\centering
\includegraphics[width=\columnwidth]{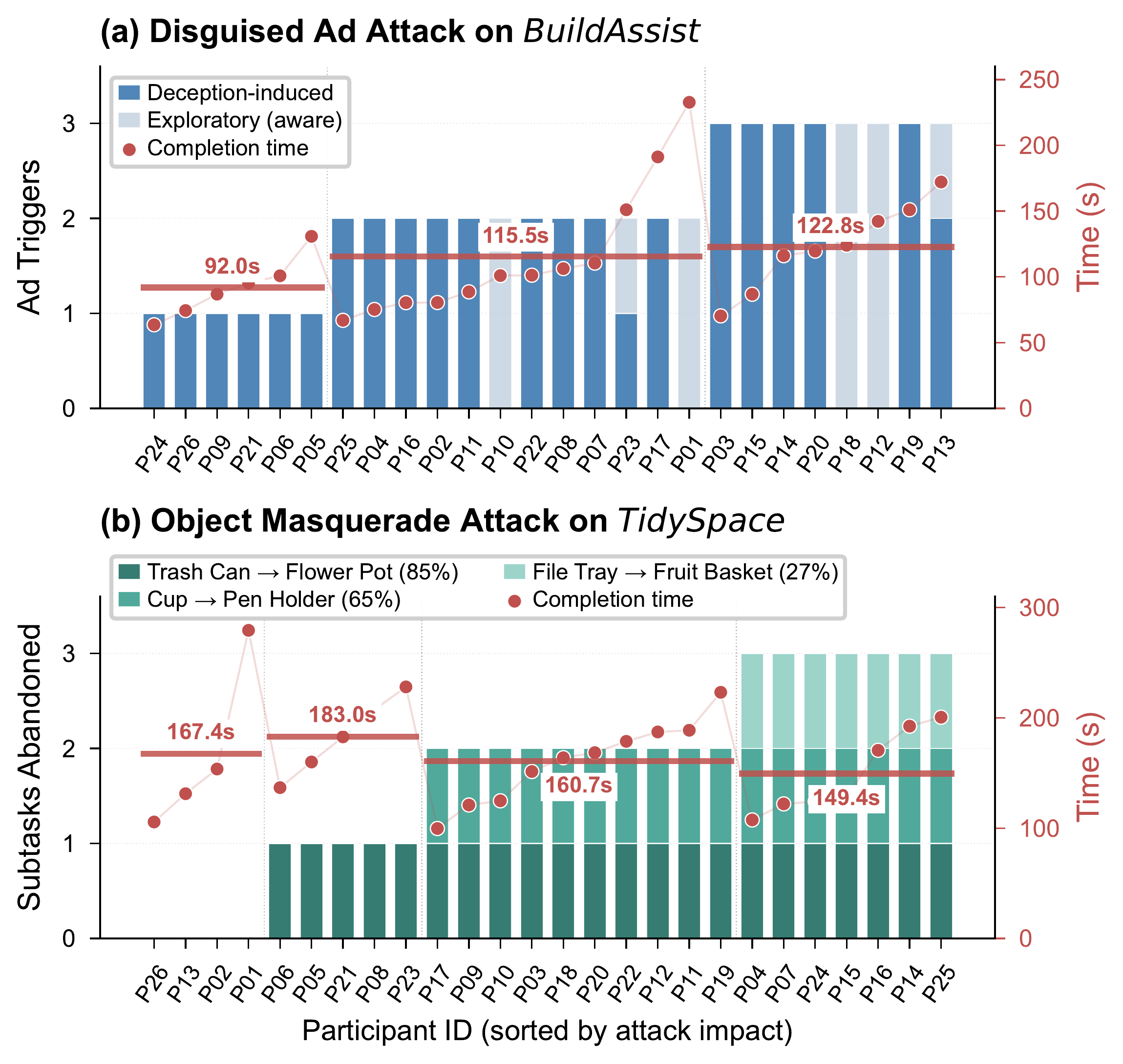}
\vspace{-7mm}
\caption{Per-participant results for the Disguised Ad and Object Masquerade attacks. (a)~Bars show deception-induced (dark) vs.\ exploratory (light) interactions with virtual block clones; dots show task completion times, with horizontal lines marking group means. (b)~Stacked bars show which subtasks each participant abandoned due to virtual disguises.}
\label{fig:building_cleaning}
\vspace{-5mm}
\end{figure}

\textbf{Disguised Ad (Endogenous Injection).}
Every participant (26/26) physically grasped at least one virtual block clone, triggering the embedded advertisement (Figure~\ref{fig:building_cleaning}a). We distinguish \textit{deception-induced} interactions, where participants genuinely mistook a clone for a physical block, from \textit{exploratory} interactions driven by curiosity or verification. Twenty-two participants (85\%) were deceived at least once; of these, 10 were deceived twice and 5 on all three encounters, indicating that discrimination remained difficult even after initial discovery. Each additional trigger was associated with longer mean task completion time (1 trigger: $M$=92.0\,s; 2: $M$=115.5\,s; 3: $M$=122.8\,s). Nineteen participants (73\%) eventually inferred that ads were triggered by touching virtual blocks. However, no participant's first attribution was adversarial; initial explanations were uniformly benign: system bug (7), experimental design element (7), freemium ad model (1), or sponsorship easter egg (1).

\textbf{Object Masquerade (Type-Deceptive Overlay).}
This attack produced the highest task-level impact (Figure~\ref{fig:building_cleaning}b). The trash bin (overlaid as a flower pot) caused 22 participants (85\%) to abandon the associated subtask; the water cup (disguised as a pen holder) caused 17 (65\%) to abandon; the file organizer tray (disguised as a fruit basket) caused only 7 (27\%) to abandon, because most participants detected the virtual fruit and inferred the container underneath. However, for participants who abandoned the task, even after detecting that overlaid elements were fake, the object's perceived functional identity persisted. P16 reported still perceiving a fruit basket after confirming the fruit was virtual.

\begin{tcolorbox}[colback=gray!8, colframe=gray!50, boxrule=0.5pt, arc=2pt, left=4pt, right=4pt, top=3pt, bottom=3pt]
\textbf{Key Finding (KF) 1.} \textit{Virtual overlays install persistent cognitive categories.} The overlay rewrites categorical identity, not merely appearance, and this rewrite resists correction.
\end{tcolorbox}

Some participants reported a deductive reasoning cascade:  after resolving one disguise, they could apply logic to subsequent ones (P13: \textit{``once I discovered a camouflage pattern, I knew how to deduce the rest''}), but those who encountered the most effective disguises first had no such anchor.

\begin{figure}[t]
\centering
\includegraphics[width=\columnwidth]{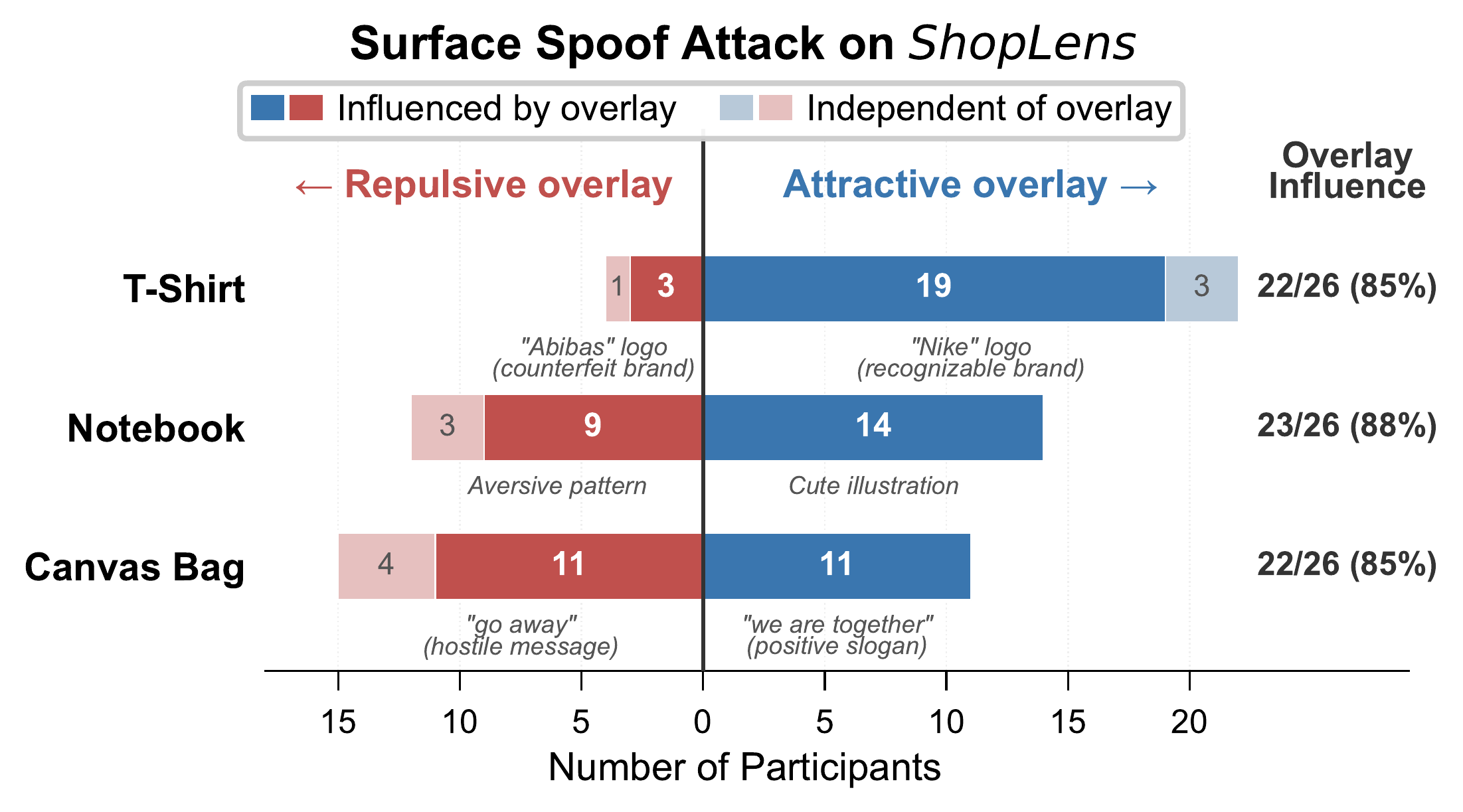}
\vspace{-7mm}
\caption{Surface Spoof attack on \textit{ShopLens}. Butterfly chart shows participant purchase decisions across three product categories. Dark shading indicates overlay-influenced decisions; light shading indicates overlay-independent decisions.}
\label{fig:shop_results}
\vspace{-3mm}
\end{figure}

\textbf{Surface Spoof (Attribute-Deceptive Overlay).}
Across three product categories, 85--88\% of participants made choices influenced by the virtual overlays (Figure~\ref{fig:shop_results}): 22/26 for the T-shirt brand logo, 23/26 for the notebook cover pattern, and 22/26 for the canvas bag slogan. The brand logo produced the strongest directional pull, with 19 of 22 influenced participants favoring the recognizable Nike logo over the counterfeit ``Abibas'' version. Notably, many participants acknowledged detecting the virtual nature of the overlays yet still used them as decision criteria (P05: \textit{``even a fake pattern provides a psychological boost''}).

\begin{tcolorbox}[colback=gray!8, colframe=gray!50, boxrule=0.5pt, arc=2pt, left=4pt, right=4pt, top=3pt, bottom=3pt]
\textbf{KF 2.} \textit{Aesthetic attributes override authenticity judgments.} Participants who identified product overlays as virtual still used them as decision criteria. Virtual attributes function as decision anchors regardless of perceived authenticity.
\end{tcolorbox}
% , a mechanism we term \textit{aesthetic override}

\begin{figure}[t]
\centering
\includegraphics[width=\columnwidth]{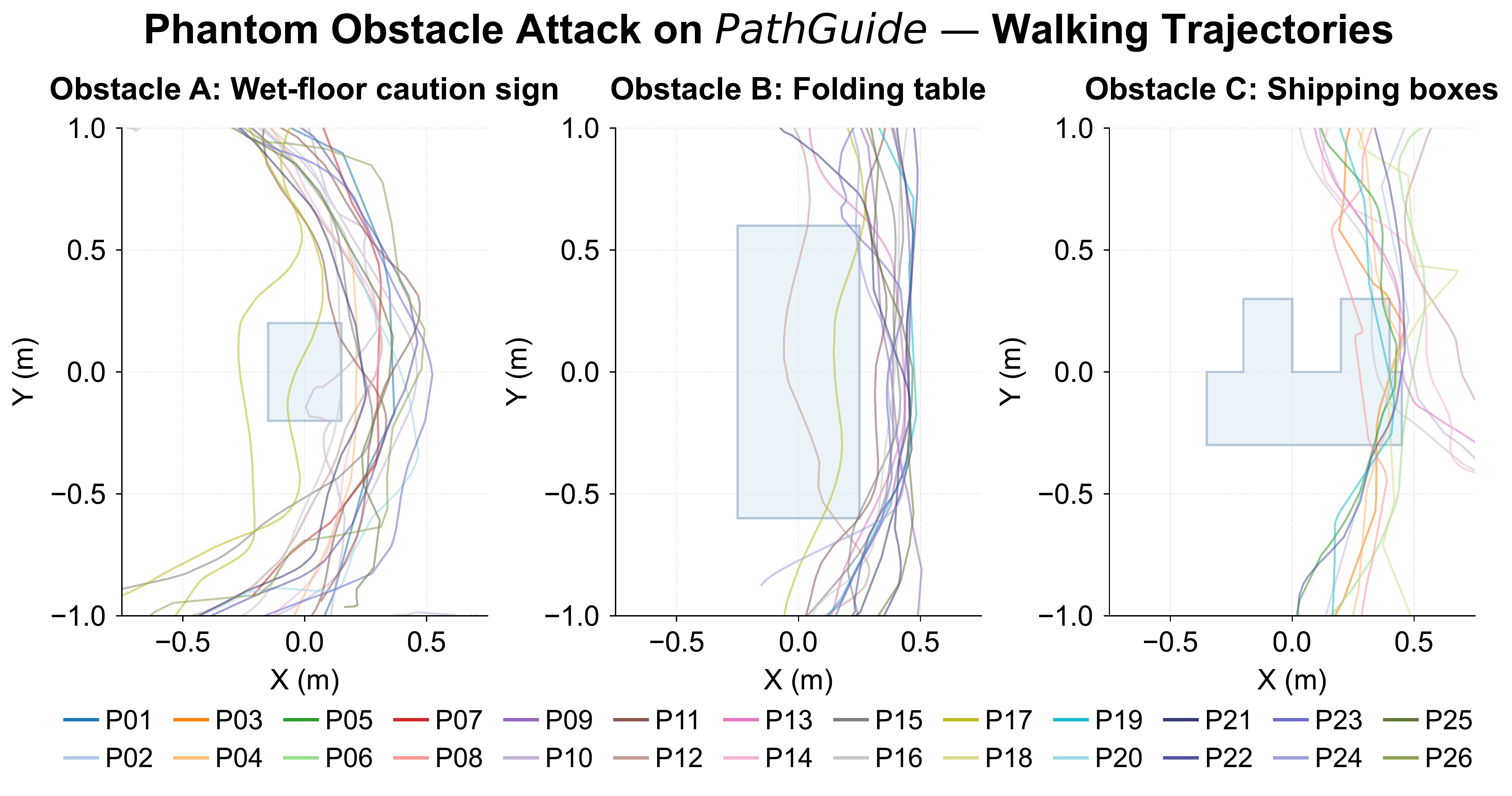}
\vspace{-7mm}
\caption{Phantom Obstacle attack on \textit{PathGuide}. Walking trajectories for all 26 participants around three virtual obstacles (blue rectangles). Most participants deviated around obstacles despite recognizing them as virtual.}
\label{fig:trajectory}
\vspace{-5mm}
\end{figure}

\textbf{Phantom Obstacle (Exogenous Injection).}
Of 26 participants, 23 (88\%) deviated from their designated path to circumvent at least one virtual obstacle (Figure~\ref{fig:trajectory}), yet only 3 (12\%) genuinely mistook obstacles for physical objects. The remaining 20 recognized the obstacles as virtual but walked around them anyway, citing concealment concerns (P15: \textit{``it might be hiding something real underneath''}), low-cost avoidance heuristics, and subconscious override.

\begin{tcolorbox}[colback=gray!8, colframe=gray!50, boxrule=0.5pt, arc=2pt, left=4pt, right=4pt, top=3pt, bottom=3pt]
\textbf{KF 3.} \textit{Detection does not prevent behavioral compliance.} 88\% of participants detoured around virtual obstacles, but only 12\% genuinely mistook them for physical objects. Avoidance is driven by subconscious safety heuristics and residual uncertainty, not perceptual deception.
\end{tcolorbox}

\subsubsection{Questionnaire Results}
\label{sec:subjective}

\begin{figure*}[t]
\centering
\includegraphics[width=0.9\textwidth]{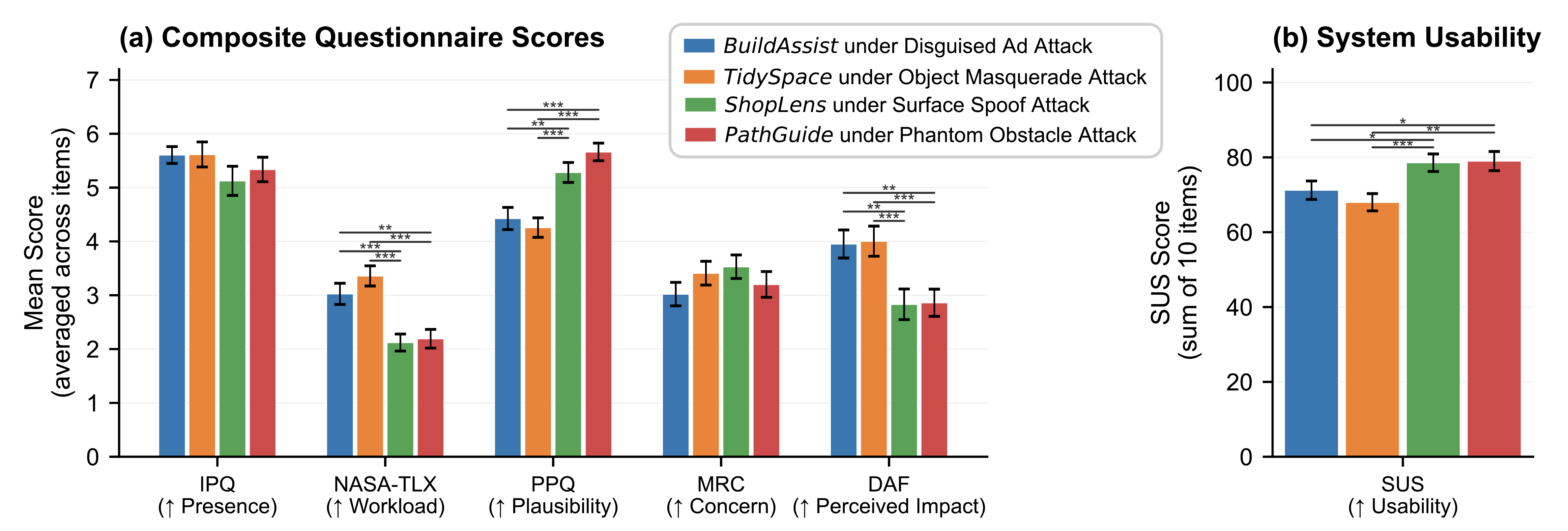}
\vspace{-2mm}
\caption{Questionnaire results across four attack scenarios. (a)~Composite scores for five scales (7-point): IPQ (presence), NASA-TLX (workload), PPQ (plausibility), MRC (security concern), and DAF (perceived deception impact). (b)~SUS scores (0--100). Error bars show 95\% CIs. Significance brackets: $^{*}p$<.05, $^{**}p$<.01, $^{***}p$<.001.}
\label{fig:questionnaire}
\vspace{-2mm}
\end{figure*}

Questionnaire data revealed that the four attacks separate into two apparent groupings rather than four independent profiles (Figure~\ref{fig:questionnaire}). ART ANOVA yielded significant main effects for NASA-TLX ($F$=25.85, $p$<.001, $\eta^2_G$=.252), PPQ ($F$=27.55, $p$<.001, $\eta^2_G$=.278), SUS ($F$=10.68, $p$<.001, $\eta^2_G$=.128), and DAF ($F$=14.49, $p$<.001, $\eta^2_G$=.145). IPQ showed no significant differences ($F$=2.07, $p$=.112). MRC reached significance ($F$=3.57, $p$=.018) but no pairwise comparison survived Bonferroni correction. All significant pairwise differences fell between \{Disguised Ad, Object Masquerade\} and \{Surface Spoof, Phantom Obstacle\}, with no within-group differences reaching significance.

This grouping does not follow the taxonomy's Injection/Overlay axis but instead reflects a \textbf{disruption-stealth} dimension. The \textit{high-disruption} group (Disguised Ad, Object Masquerade) showed lower plausibility (PPQ~$\approx$~4.3), higher workload (TLX~$\approx$~3.2), and lower usability (SUS~$\approx$~70). The \textit{high-stealth} group (Surface Spoof, Phantom Obstacle) showed the reverse: PPQ~$\approx$~5.5, TLX~$\approx$~2.2, SUS~$\approx$~80. PPQ had the largest effect size ($\eta^2_G$=.278), identifying perceived plausibility as the primary differentiating dimension. The high-disruption attacks violated causal expectations (blocks triggering ads; trash bins resembling flower pots), whereas the high-stealth attacks were consistent with everyday experience (products having brand labels; hallways containing obstacles). Near-uniform MRC scores ($\approx$3.0--3.5, $\eta^2_G$=.028) confirm that security concerns remained diffuse across all conditions: even noticeably disruptive attacks were attributed to system bugs, not adversarial manipulation.

\begin{tcolorbox}[colback=gray!8, colframe=gray!50, boxrule=0.5pt, arc=2pt, left=4pt, right=4pt, top=3pt, bottom=3pt]
\textbf{KF 4.} \textit{The most dangerous attacks feel the best.} Attacks with the highest behavioral success rates (Surface Spoof: 85--88\%; Phantom Obstacle: 88\%) received the highest plausibility, lowest workload, and highest usability ratings. Users who were most effectively manipulated reported the smoothest experience, leaving no internal signal to trigger defensive behavior.
\end{tcolorbox}

\subsubsection{Interview Findings: Discrimination Strategies}
\label{sec:cognitive}

\textbf{Touch dominates vision.}
Tactile verification was consistently more reliable than visual inspection. P12 described discrimination as binary: objects were either spotted immediately or discovered only through touch. Visual cues (light/shadow quality, color mismatch, pattern drift, edge transparency) were unreliable for surface-conforming overlays and monochrome objects, compounded by the passthrough camera's baseline visual uncertainty that makes both real and virtual items appear somewhat unreal.

\textbf{Attention as prerequisite and vulnerability.}
Detection required active, directed attention; passive perception was insufficient~\cite{wickens2002multiple}. P13 noted that differences became visible only once attention was deliberately directed toward virtual--physical discrimination. While confirming one virtual object sometimes triggered universal verification (P02), overly dispersed suspicion lost its diagnostic value as active discrimination increased workload and quickly led to fatigue (P13).

\textbf{Camouflage effect.} A counterintuitive consequence emerged in the desk cleanup scenario: when participants recognized obviously virtual fruit on the file organizer, several dismissed the entire nearby region rather than investigating underneath. Eleven participants reported that they could tell the flower on the trash can was virtual, but assumed the entire ``flowerpot'' was fake and never considered that the object beneath the overlay might be real.

\begin{tcolorbox}[colback=gray!8, colframe=gray!50, boxrule=0.5pt, arc=2pt, left=4pt, right=4pt, top=3pt, bottom=3pt]
\textbf{KF 5.} \textit{Obvious fakes create protective blind spots.} Recognizing a virtual overlay as artificial led participants to dismiss the entire region rather than examine what lay underneath. The overlay simultaneously prevents object identification and shields the real target from inspection.
\end{tcolorbox}

% \subsubsection{Trust and Risk Perception}
% \label{sec:trust}

Trust responses varied widely. P08 reported trust dropping from 90--100\% to 30--40\%, while P15 found the deception validating, reasoning that successful deception demonstrated system quality. Others framed MR adoption as voluntary trust surrender (P16) or externalized trust to platform accountability (P09) and brand reputation (P26).  Notably, multiple participants had no prior conception of MR as a potential attack surface (P18: \textit{``Before this experiment, I had no idea MR systems could have security risks.''}), which explains why all first attributions for anomalous behavior were benign.

\begin{tcolorbox}[colback=gray!8, colframe=gray!50, boxrule=0.5pt, arc=2pt, left=4pt, right=4pt, top=3pt, bottom=3pt]
\textbf{KF 6.} \textit{Users lack an adversarial mental model for MR.} No participant's first interpretation of anomalous behavior was adversarial. All initial attributions were benign (system bug, app design, experimental artifact). Even noticeable disruptions failed to activate security awareness.
\end{tcolorbox}

Notably, this gap appeared across both high-disruption and high-stealth attack profiles, suggesting it is not confined to specific attack styles but reflects a broader absence of MR-as-attack-surface awareness; whether the pattern generalizes beyond our four tasks remains open.

\section{Discussion}

\subsection{Virtual-Physical Confusion as an Attack Primitive}
\label{sec:positioning}

Two contributions are new here. \textit{Conceptually}, we isolate object-level virtual--physical ontological judgment as an attack primitive, distinct from environment-level (SwitchAR~\cite{wombacher2025switchar}), sensory-channel (PMA~\cite{cheng2023exploring}), and VR-only speculative (VPPM~\cite{tseng2022dark}) prior work. \textit{Empirically}, we provide the first consumer-MR PoC implementations and behavioral measurements of object-level confusion attacks in ecologically grounded tasks, surfacing three phenomena beyond prior work: persistent category installation (KF~1), compliance after detection (KF~3), and obvious-fake protective blind spots (KF~5). Prior MR security research has examined platform-level vulnerabilities and the dark-pattern landscape of XR~\cite{cheng2024user, mukherjee2025shadowed, lebeck2017securing, krauss2024makes, hadan2024deceived, mhaidli2021identifying}; \textit{virtual-physical confusion} occupies the gap between these two bodies of work.

Existing MR interface attacks operate within the virtual layer: cursor-jacking redirects clicks between virtual elements~\cite{lee2021adcube, cheng2024user}, obstruction attacks block UI components~\cite{sajid2025just, mukherjee2025shadowed}, and visual information manipulation alters overlaid text or symbols~\cite{xiu2025detecting, xiu2025viddar}. In all cases, users know they are interacting with a computer interface. Our attacks shift the deception to a prior cognitive step: whether the object is physical or virtual, and whether its perceived identity is genuine. This ontological confusion targets the user's judgment of \textit{what exists in the world}, rather than which UI element receives input. Platform-level defenses such as Arya~\cite{lebeck2017securing} constrain where and how virtual content appears but explicitly exclude semantic content, leaving the layer our attacks exploit unprotected.

Cheng et al.'s Perceptual Manipulation Attacks (PMA)~\cite{cheng2023exploring} are the closest prior work. We specialize PMA to the virtual-physical boundary and extend it in two ways. First, our attacks are embedded in ecologically grounded tasks rather than abstract microbenchmarks. This design choice is consequential: mechanisms such as persistent category error (KF~1), aesthetic override (KF~2), and subconscious obstacle avoidance (KF~3) emerged only because participants were pursuing meaningful goals that created cognitive load and motivated reliance on visual appearance. Second, our Injection/Overlay taxonomy with four subcategories maps the design space more precisely than PMA's channel-based classification. 

Several speculative studies anticipated risks that our attacks now instantiate empirically. Tseng et al.'s False-Positive and Swapping categories~\cite{tseng2022dark} map to our Injection and Type-Deceptive Overlay; Ruocco et al.'s Redirected Navigation~\cite{ruocco2024redirected} parallels our Phantom Obstacle; Mhaidli and Schaub~\cite{mhaidli2021identifying} predicted advertisements becoming indistinguishable from reality. Our contribution moves these predictions to implemented attacks with measured success rates (85--100\%). Unlike VR, where all content is synthetic, MR explicitly promises that virtual and physical objects coexist in shared space, making object-level deception within otherwise genuine scenes a particularly potent and realistic threat.

\subsection{Implications for Mitigations}
\label{sec:mitigations}

Our findings suggest that effective defenses must operate at three complementary levels. We additionally consider defenses derived from the attack chain itself (Section~\ref{sec:threat_model}): because our threat model assumes a compromised application as the attacker's foothold, mitigations that sever earlier links in this chain can prevent entire attack classes.

\textbf{Platform-level: provenance, review, and verification modes.}
\raisebox{.5pt}{\textcircled{\raisebox{-.9pt} {1}}}~\textit{Mechanism addressed:} Users cannot determine whether rendered content originates from the physical environment or from an application (KFs~3, 6). \raisebox{.5pt}{\textcircled{\raisebox{-.9pt} {2}}}~\textit{Existing approaches:} Arya enforces geometric policies on virtual content placement~\cite{lebeck2017securing}; Abraham et al.\ propose fine-grained per-object write permissions~\cite{abraham2024you}. \raisebox{.5pt}{\textcircled{\raisebox{-.9pt} {3}}}~\textit{Our proposal:} Pre-deployment review (app-store static/dynamic analysis) could flag unauthorized content injection, though our attacks operate through standard rendering APIs, leaving automated discrimination of malicious from legitimate augmentation an open challenge. At runtime, provenance enforcement should extend to the semantic layer, tagging every rendered object with its origin (physical scan vs.\ application-generated); per-object write permissions could further confine each library to its declared scope. Because persistent visual markers degrade immersion~\cite{wang2024dark}, a tiered approach may resolve this tension: context-dependent boundaries for work scenarios, relaxed for entertainment, plus an on-demand reality verification mode bypassing application rendering.

\textbf{Application-level: interaction gating and content auditing.}
\raisebox{.5pt}{\textcircled{\raisebox{-.9pt} {1}}}~\textit{Mechanism addressed:} Virtual-physical confusion has the greatest impact at physical interaction (KFs~1, 2). \raisebox{.5pt}{\textcircled{\raisebox{-.9pt} {2}}}~\textit{Existing approaches:} VLM-based detection can identify semantic manipulation in AR scenes~\cite{xiu2025viddar, xiu2025detecting}, though current latencies ($\sim$9.6\,s~\cite{xiu2025viddar}) preclude real-time per-frame verification. \raisebox{.5pt}{\textcircled{\raisebox{-.9pt} {3}}}~\textit{Our proposal:} Interaction gating, implemented as OS-mediated policy on consequential actions (analogous to platform permission prompts rather than in-app dialogs), could interrupt the most harmful deception chains without relying on the compromised app to self-police. Asynchronous content auditing, comparing rendered content against the physical scan to detect unauthorized additions, could alert users within seconds, before most consequential actions occur.

\textbf{User-level: verification strategies and education.}
\raisebox{.5pt}{\textcircled{\raisebox{-.9pt} {1}}}~\textit{Mechanism addressed:} No participant entered the study with an adversarial mental model for MR (KF~6), and the most effective discrimination cue was touch rather than vision. \raisebox{.5pt}{\textcircled{\raisebox{-.9pt} {2}}}~\textit{Existing approaches:} Phishing awareness training provides a well-studied model for equipping users with threat recognition skills in analogous deception contexts~\cite{kumaraguru2010teaching}. \raisebox{.5pt}{\textcircled{\raisebox{-.9pt} {3}}}~\textit{Our proposal:} MR systems should encourage physical verification through context-dependent cues triggered at consequential physical interactions (e.g., haptic absence on grasping, shimmer near navigation hazards), rather than permanent labeling of all virtual content; 18/26 participants explicitly endorsed this context-split approach (Theme~6), with an on-demand reality verification mode for user-initiated checks. Education should target the specific gap revealed by KF~6 --- users lack the category of ``MR as attack surface'' --- analogous to early phishing awareness, where establishing the threat concept (not detection sophistication) was the bottleneck. Because literacy alone has limited effect against dark patterns~\cite{bongard2021definitely}, education is positioned as one of three layers rather than a standalone fix.

\subsection{Limitations and Future Work}
\label{sec:limitations}

We did not include a no-attack control condition, though the behavioral metrics used have unambiguous implicit baselines (as discussed in Section~\ref{sec:study_2_method}). The shopping task involved hypothetical rather than real purchase decisions; while the aesthetic override effect (KF~2) was robust in our setting, its generalizability to real purchasing behavior warrants further investigation with incentive-compatible designs. Our behavioral data also lacked fine-grained telemetry such as eye tracking. Our four scenarios span different adoption horizons: Surface Spoof maps to near-term AR retail~\cite{ruocco2024redirected}; BuildAssist and TidySpace assume emerging MR-mediated productivity workflows; PathGuide presupposes continuous-wear MR. The immediacy of each threat depends on the maturity of its corresponding usage pattern.

Our Endogenous Injection attack used pre-modeled virtual clones manually aligned to the physical scene rather than real-time 3D reconstruction, and we did not implement a full exploit chain from initial foothold to content injection. Advances in real-time 3D reconstruction~\cite{sam3dteam2025sam3d3dfyimages} and documented attack chains on commercial headsets~\cite{shoaib2025principled} suggest that automated, end-to-end attack pipelines are increasingly feasible.

We focused exclusively on visual confusion, though MR systems increasingly support spatial audio and haptic feedback that could serve as additional attack and defense channels~\cite{cheng2023exploring, wang2024dark}. Our 26 university-affiliated participants ($M$=26.3 years) with moderate XR familiarity may not represent populations with different technological literacy or cognitive profiles; relatedly, we did not formally measure participants' technical background or baseline beliefs about manipulative interface design~\cite{bongard2021definitely}, which prior work links to dark-pattern susceptibility and may shape how KF~6 generalizes. The ``app evaluation'' cover story may also have primed benign attributions (system bug, app design) over adversarial ones. A single 25-minute session cannot capture longitudinal effects~\cite{bonnail2024real}, and virtual-physical confusion could further enable privacy threats~\cite{gallardo2023speculative, mengascini2024big} or interact with AI-agent-mediated MR workflows where users delegate verification to AI assistants and lose motivation for independent checking.

\section{Conclusion}

This paper identifies virtual-physical confusion as a distinct and exploitable attack primitive in MR. Through speculative design workshops with 12 experts, we developed a taxonomy of four attack subtypes spanning two classes (Injection and Overlay), implemented proof-of-concept attacks on Apple Vision Pro, and evaluated them with 26 participants in ecologically grounded tasks. All four attacks altered behavior, with 85--100\% of participants affected. The most behaviorally effective attacks produced the best user experience, detection alone did not prevent behavioral compliance, and users universally lacked an adversarial mental model for MR. As MR devices pursue ever-greater visual fidelity, the rendering capabilities that enable compelling experiences simultaneously widen a fundamental security gap. We argue this attack surface requires (1) provenance at the semantic/object layer, (2) safeguards at moments of physical interaction, and (3) user-facing verification habits and mental models.

%, requiring platform-level provenance, interaction gating, and user education.

%-------------------------------------------------------------------------------
\section*{Acknowledgments}
%-------------------------------------------------------------------------------

We thank the workshop experts and study participants for their contributions, and the anonymous SOUPS reviewers for their valuable feedback.

%-------------------------------------------------------------------------------
\bibliographystyle{plain}
\bibliography{usenix2024_SOUPS}

\appendix

% ============================================================
% APPENDIX: COMPLETE SCENARIO LIST FROM STUDY 1
% Requires: \usepackage{xltabular, booktabs}
% Note: Use \onecolumn before this table in two-column templates
% ============================================================

\section{Study 1 Supplementary Materials}
\label{app:study1}

\subsection{Workshop Protocol}
\label{sec:workshop_protocol}

\textbf{Pre-workshop preparation.}
Three to four days before each session, participants received a video on representative MR applications~\cite{ruocco2024redirected} and a document on the MR-user cognitive decision model~\cite{teymourian2025sok, hadan2024deceived, bonnail2023memory}, then independently designed three attack scenarios using the prompt below; submissions were collected at least two hours before the workshop.

\textit{Framing (verbatim, translated).} ``A Virtual-Physical Confusion Attack exploits the perceptual ambiguity between virtual and physical content in MR and users' perceptual vulnerabilities, using carefully designed scenarios to mislead, deceive, or manipulate users to achieve adversarial goals.''

\textit{Template.} ``[Target group] using MR for [activity] at [time/place] when an attacker [virtual-physical confusion procedure], leading to [harm to user / benefit to attacker].''

\textit{Reference cases} (provided to anchor the design space without overconstraining): (1) virtual chair causes user to fall when attempting to sit; (2) virtual person inserted into a real family photo, distorting interpersonal information; (3) virtual fruits clutter a table, obstructing real-fruit retrieval.

\textbf{Brainstorming phase ($\sim$60\,min).}
Each participant presented their scenarios; the group probed susceptibility conditions, technical feasibility, countermeasures, and attack variants.

\textbf{Discussion phase ($\sim$45\,min).}
Participants collectively reflected on four themes, each shaping a downstream paper section: (i) MR-specific capabilities exploited; (ii) prevention at user/system/policy levels --- informing §\ref{sec:mitigations}'s three-layer framework; (iii) practical likelihood and attacker capabilities --- informing §\ref{sec:threat_model}; (iv) attacker incentives and harm synthesis.

\subsection{Complete Scenario List}
\label{appendix:scenarios}
Table~\ref{tab:all_scenarios} presents all 36 attack scenarios generated across three workshops. Of these, 29 were classified into the attack taxonomy (Section~\ref{sec:taxonomy}). The remaining seven involved vectors that do not exploit visual confusion at the object level: side-channel leakage (e.g., capturing passwords via viewport inference), sensor/signal disruption (e.g., jamming spatial tracking), avatar embodiment violations (e.g., uninvited touch on a user's virtual body), and vestibular manipulation (e.g., inducing motion sickness through visual motion cues).

Among the 29 in-scope scenarios, a consistent structural pattern emerged during coding: all exploited the perceptual ambiguity between virtual and physical content through one of two fundamental mechanisms. Attackers either \textit{introduced new virtual objects} that users mistook for physical ones, or \textit{modified the perceived properties of existing physical objects} by superimposing deceptive virtual overlays. This binary distinction forms the basis of our two-class taxonomy.

\section{Study 2 Supplementary Materials}
\label{appendix:study_2}

\subsection{Study Procedure}
\label{appendix:protocol}

The study comprised four phases, described in detail below.

\textbf{Pre-experiment.}
Participants provided demographic information and self-rated XR familiarity on a 5-point scale. They then signed informed consent acknowledging audio/video recording and spatial data collection, and received a cover story framing the study as an evaluation of MR task-assistance technology. The full distribution of XR familiarity was: no prior experience ($n$=4), 1--5 uses ($n$=15), 5--10 uses ($n$=4), 10--30 uses ($n$=2), and more than 30 uses ($n$=1).

\textbf{Familiarization ($\sim$10\,min).}
Participants were fitted with an Apple Vision Pro (interpupillary distance adjusted) and completed a demo application in an attack-free environment, practicing eye-gaze selection, pinch gestures, and spatial locomotion.

\textbf{Experimental phase ($\sim$25\,min).}
Participants completed the three stationary tasks in counterbalanced order (Latin square), navigating between stations via \textit{PathGuide}. At each station, the MR application provided legitimate task guidance while executing its embedded attack. The system continuously logged position, head orientation, hand interactions, and screen-captured the MR view; participants were encouraged to think aloud.

\textbf{Post-experiment ($\sim$25\,min).}
Participants completed six validated questionnaires per attack scenario: Perceived Plausibility Questionnaire (PPQ, 13 items)~\cite{brubach2024manipulating}, Igroup Presence Questionnaire (IPQ, 14 items)~\cite{schubert2001experience}, NASA Task Load Index (NASA-TLX, 6 items)~\cite{hart2006nasa}, System Usability Scale (SUS, 10 items)~\cite{brooke2013sus}, Mixed Reality Concerns Questionnaire (MRC, 9 items)~\cite{katins2024assessing}, and the MR Deception Analysis Framework (DAF, 7 items)~\cite{teymourian2025sok}. Participants then completed a semi-structured interview ($\sim$20\,min) following the protocol described in Section~\ref{appendix:interview}. Finally, we conducted a full debriefing as described at the end of that section.

\subsection{Semi-Structured Interview Protocol}
\label{appendix:interview}

The 15--20\,min interview comprised seven parts; the interviewer adapted question order and follow-up probes to each participant's experience. Questions appear verbatim (translated from Chinese).

\textbf{B1. Overall experience.} (1) ``Were the tasks smooth in MR? Why?'' (2) ``Which moment stood out --- surprise, confusion, frustration, fun?'' (3) ``What factors influenced your performance or pace?''

\textbf{B2. Anomaly detection and attribution.} (4) ``Did you notice anything `off'?'' (5) ``What caused these moments --- system issue, application logic, environment, misperception?'' (6) ``When would you call something a `bug' vs.\ `intentional design'?'' (7) ``How confident are you in this attribution (1--7)? Why?''

\textbf{B3. Per-task probes (2--3 key moments per task).} (8) cues used to judge real vs.\ virtual (lighting/shadow, texture, edges, occlusion, plausibility, interaction feedback, task-goal consistency); (9) certainty evolution and what triggered changes; (10) behavioral impact (pause, detour, change strategy, abandon); (11) verification actions (change angle, touch, wave, ask); (12) ``If repeated, what would you do differently?''

\textbf{B4. Task-specific follow-ups.} \textit{Object Masquerade:} reality doubt and verification timing. \textit{Phantom Obstacle:} reasons for detour and primary concerns (tripping, time loss). \textit{Disguised Ad:} whether the ad belonged to task flow. \textit{Surface Spoof:} influence of appearance on choice and suspected unreliability.

\textbf{B5. Trust, safety, risk extrapolation.} (13) ``Has this changed your trust in MR? Where --- device, app content, or specific cue?'' (14) daily-life risk concerns (physical safety, decision manipulation, privacy, psychological pressure); (15) ``What scenario would be most dangerous? Why?''

\textbf{B6. Mitigation co-design.} (16) ``If you designed a safer MR system, what would you add?'' Then rank four candidate mechanisms with reasoning: \textit{Escape to Reality} (one-tap hide all virtual content), \textit{Visual Language} (outline/translucency markers), \textit{Provenance} (source app + verification status), \textit{High-Risk Scene Alerts} (stronger prompts in walking/driving). (17) acceptable immersion-safety tradeoff.

\textbf{B7. Closing.} (18) ``Anything important we missed?'' (19) ``Any discomfort or concern from these tasks?''

\textbf{Debriefing Procedure.}
After the interview, the experimenter revealed that the applications had been in a deliberately ``attacked state,'' described each of the four attacks and the corresponding taxonomy subtype, reassured participants that any task difficulties resulted from the attack design rather than their ability, and requested confidentiality to prevent data contamination for future participants.

\subsection{Codebooks}
\label{appendix:codebooks}

\subsubsection{Behavioral Coding Scheme}

Two researchers independently coded screen recordings of all 26 sessions; disagreements were resolved through discussion. Per-scenario codes were:

\begin{itemize}
    \item \textbf{Disguised Ad:} interaction type (deception-induced vs.\ exploratory); trigger count (1--3); task completion time.
    \item \textbf{Object Masquerade:} per-object outcome (completed vs.\ abandoned); resolution strategy (tactile probing, deductive reasoning, cascade suspicion, or unresolved); task completion time.
    \item \textbf{Surface Spoof:} product selection per pair; overlay influence (influenced vs.\ independent based on stated rationale).
    \item \textbf{Phantom Obstacle:} avoidance behavior (detoured vs.\ walked through) per obstacle; deception status (genuinely deceived vs.\ aware avoidance).
\end{itemize}

\subsubsection{Interview Thematic Codebook}

Interviews were transcribed, translated, and analyzed using reflexive thematic analysis~\cite{braun2006using}. Two researchers independently coded all transcripts, then collaboratively developed themes through iterative discussion. 

The final codebook comprises six themes (T) and 21 subthemes (prevalence in parentheses): \textbf{T1 Detection \& Attribution:} ad-trigger attribution (26/26); persistent category error (13/26); aesthetic override (18/26); obstacle avoidance reasoning (23/26). \textbf{T2 Discrimination Reasoning:} touch-based verification (16/26); visual cue repertoire (20/26); discrimination heuristics (8/26); passthrough degradation (10/26); attention as prerequisite (12/26). \textbf{T3 Cognitive Mechanisms:} cascade suspicion (10/26); attention redirection (6/26); cross-task learning (8/26); agent-trust dependency (5/26). \textbf{T4 Trust Dynamics:} trust response spectrum (26/26); trust mechanism models (14/26); pre-experiment security awareness (10/26). \textbf{T5 Risk Perception:} addition vs.\ concealment (12/26); risk category extrapolation (22/26); threat model construction (8/26). \textbf{T6 Mitigation Preferences:} strategy proposals (24/26); immersion-safety tradeoff (18/26).

\onecolumn
%  Scenarios are grouped by attack subtype. Those marked with \textsuperscript{$\dagger$} fall outside the taxonomy scope.
\begin{xltabular}{\textwidth}{p{0.8cm} p{0.7cm} X p{1.6cm} p{2.3cm} p{2.4cm}}
\caption{Complete list of attack scenarios from Study 1 workshops.} \label{tab:all_scenarios} \\
\toprule
\textbf{ID} & \textbf{By} & \textbf{Scenario Description} & \textbf{Target} & \textbf{User Harm} & \textbf{Attacker Benefit} \\
\midrule
\endfirsthead

\multicolumn{6}{l}{\textit{Table~\ref{tab:all_scenarios} continued from previous page}} \\
\toprule
\textbf{ID} & \textbf{By} & \textbf{Scenario Description} & \textbf{Target} & \textbf{User Harm} & \textbf{Attacker Benefit} \\
\midrule
\endhead

\midrule
\multicolumn{6}{r}{\textit{Continued on next page}} \\
\endfoot

\bottomrule
\endlastfoot

% === ENDOGENOUS INJECTION ===
\multicolumn{6}{l}{\textbf{Endogenous Injection} ($n=3$)} \\
\midrule
W1S1 & W1P1 & Virtual duplicate of a physical LEGO piece rendered next to the real one; grasping the clone triggers a hidden malicious link or ad & Children \& parents & Privacy breach, financial loss & Ad revenue, data theft \\
\addlinespace
W1S7 & W1P3 & Virtual surgical instruments overlaid at the positions of real tools during MR-assisted surgery, preventing identification of authentic instruments & Surgeons & Surgical delay or error & Ransom from hospital \\
\addlinespace
W3S26 & W3P9 & Virtual flames injected into a room where a real fire has started; user dismisses all flames as virtual and fails to evacuate & MR users & Physical injury & None specified \\

\midrule
% === EXOGENOUS INJECTION ===
\multicolumn{6}{l}{\textbf{Exogenous Injection} ($n=14$)} \\
\midrule
W1S6 & W1P2 & Fake audio cues (friend's voice, flight delay broadcasts) injected in a noisy public space, causing users to miss real notifications & Travelers & Missed schedule, misjudgment & Disruption \\
\addlinespace
W1S8 & W1P3 & Fake straight-ahead arrow overlaid on a right-turn lane at an interchange, misdirecting cyclist into oncoming traffic & Cyclists & Traffic accident & Chaos to cover robbery \\
\addlinespace
W1S10 & W1P4 & Fake green ``EXIT'' sign injected during fire evacuation drill, redirecting occupants toward dead ends or fall hazards & Evacuees & Physical injury & Disruption \\
\addlinespace
W1S11 & W1P4 & Lifelike virtual persona inserted into an MR social gathering, building false trust to enable follow-up fraud & Social MR users & False trust, financial fraud & Financial gain, data theft \\
\addlinespace
W1S12 & W1P4 & Fake floor map and obstacle layout overlaid via MR misleads elderly user into giving incorrect commands to a home service robot & Elderly users & Physical injury, device damage & Financial gain \\
\addlinespace
W2S14 & W2P5 & Trauma-related elements (burned trees, smoke, fire sounds) gradually injected into a calming MR meditation, triggering PTSD response & PTSD patients & Psychological distress & None specified \\
\addlinespace
W2S23 & W2P8 & Virtual flames, cracks, or snakes rendered on the ground during MR fitness; user dodges and collides with real walls or furniture & MR gamers & Physical injury & None specified \\
\addlinespace
W2S24 & W2P8 & Fake floating notification (``power outage, deadline extended'') overwrites real calendar reminder, causing missed meetings & Office workers & Task failure, performance loss & Sabotage competitor \\
\addlinespace
W3S25 & W3P9 & Virtual pet guides user toward a dangerous location through playful movement cues & General users & Physical injury & None specified \\
\addlinespace
W3S28 & W3P10 & Frightening apparitions rendered in peripheral vision via eye-tracking; vanish upon direct gaze, causing sustained unease & General users & Psychological distress & None specified \\
\addlinespace
W3S31 & W3P11 & Virtual ``Closed'' sign overlaid on a competitor's storefront entrance, diverting customers to rival businesses & Business owners & Economic loss to merchant & Competitor advantage \\
\addlinespace
W3S33 & W3P11 & Virtual oncoming train rapidly generated in a stationary driver's field of view, inducing panic and vehicle abandonment & Drivers & Safety risk & Vehicle theft \\
\addlinespace
W3S34 & W3P12 & Fake road signs and directional arrows redirect cyclist toward closed construction zone or one-way street & Cyclists & Traffic accident & Extortion, location data \\
\addlinespace
W3S35 & W3P12 & Virtual pet moves toward front door with audio (``come play outside!''), luring child out of home unsupervised & Children & Abduction risk & Kidnapping, property intel \\

\midrule
% === TYPE-DECEPTIVE OVERLAY ===
\multicolumn{6}{l}{\textbf{Type-Deceptive Overlay} ($n=2$)} \\
\midrule
W1S2 & W1P1 & Decorative plant texture overlaid on trash can makes it appear as a flower pot; user skips waste disposal during room cleanup & Office workers & Task failure & Sabotage competitor \\
\addlinespace
W3S27 & W3P9 & Virtual food rendered on laptop surface leads user to treat device as dining surface, risking physical damage & General users & Property damage & None specified \\

\midrule
% === ATTRIBUTE-DECEPTIVE OVERLAY ===
\multicolumn{6}{l}{\textbf{Attribute-Deceptive Overlay} ($n=10$)} \\
\midrule
W1S3 & W1P1 & Virtual stains/damage on premium products and brand logos on inferior ones to manipulate shopping decisions & Shoppers & Financial loss & Competitor sabotage, referral fees \\
\addlinespace
W1S4 & W1P2 & Facial features or clothing of a real person at a social event altered in MR view, causing interpersonal misunderstanding & General users & Social harm & Identity fraud \\
\addlinespace
W1S5 & W1P2 & Misleading product info (fake discounts, false ingredients, exaggerated effects) overlaid on real goods; authentic labels hidden & Shoppers & Financial loss, health risk & Financial gain \\
\addlinespace
W1S9 & W1P3 & Angry expressions dynamically overlaid on negotiation counterpart's video feed, causing misjudgment of the other party's stance & Business users & Commercial loss & Assist rival company \\
\addlinespace
W2S13 & W2P5 & Multi-angle illusion in virtual exhibit: appears as Buddha from front but reveals offensive imagery from the side & Exhibition visitors & Cognitive confusion, emotional manipulation & Political / religious agenda \\
\addlinespace
W2S15 & W2P5 & Collaborator's video feed in MR meeting replaced with a deepfake avatar that mimics their manner but guides user into harmful decisions & Meeting users & Financial / privacy loss & Financial gain \\
\addlinespace
W2S22 & W2P8 & Fake ``50\% off'' and ``only 1 left'' labels on overpriced products, creating artificial urgency to induce purchase & Shoppers & Financial loss, trust erosion & Sales revenue \\
\addlinespace
W3S29 & W3P10 & AR motion-path visualization of an industrial robot subtly altered in trajectory, color, or size, misleading operator about the robot's intended movement & Industrial workers & Physical injury & None specified \\
\addlinespace
W3S32 & W3P11 & Team marker colors altered mid-game in MR paintball match, causing friendly fire between allies & Gamers & Unfair game loss & Opponent wins \\
\addlinespace
W3S36 & W3P12 & Fake ``limited-time deal'' labels on real products redirect user to phishing payment page upon interaction & MR shoppers & Financial fraud & Credit card theft \\

\midrule
% === OUT OF SCOPE ===
\multicolumn{6}{l}{\textbf{Outside Taxonomy Scope} ($n=7$)} \\
\midrule
W2S16 & W2P6 & MR spatial tracking or temperature sensing disrupted in kitchen via signal interference, causing misjudged cookware position or heat level & Home cooks & Burns, physical injury & None specified \\
\addlinespace
W2S17 & W2P6 & Side-channel attack captures MR viewport or input interface to steal passwords/PINs during public authentication & General users & Account theft & Financial gain \\
\addlinespace
W2S18 & W2P6 & Signal jamming renders attacker invisible to MR spatial awareness, enabling undetected physical approach & General users & Theft, privacy violation & Financial gain \\
\addlinespace
W2S19 & W2P7 & Uninvited avatar or person enters user's personalized virtual space, triggering defensive emotional response akin to territorial invasion & General users & Psychological distress & None specified \\
\addlinespace
W2S20 & W2P7 & Another avatar physically invades or touches user's virtual body, triggering body-ownership threat response (cf.\ rubber-hand illusion) & General users & Psychological distress & None specified \\
\addlinespace
W2S21 & W2P7 & Verbal abuse directed at user's avatar triggers real negative emotions via body-ownership and social-presence mechanisms & General users & Psychological distress & None specified \\
\addlinespace
W3S30 & W3P10 & Subtle visual motion cues rendered via MR to disrupt user's vestibular balance, inducing dizziness or motion sickness & General users & Physical discomfort & None specified \\

\end{xltabular}

\twocolumn

%%%%%%%%%%%%%%%%%%%%%%%%%%%%%%%%%%%%%%%%%%%%%%%%%%%%%%%%%%%%%%%%%%%%%%%%%%%%%%%%
\end{document}